# Concepts and their Use for Modelling Objects and References in Programming Languages

Alexandr Savinov

Institute of Mathematics and Computer Science, Academy of Sciences of Moldova
Academiei 5, MD-2028 Chisinau, Moldova

http://www.conceptoriented.com
savinov@conceptoriented.com

In the paper a new programming construct, called concept, is introduced. Concept is pair of two classes: a reference class and an object class. Instances of the reference classes are passed-by-value and are intended to represent objects. Instances of the object class are passed-by-reference. An approach to programming where concepts are used instead of classes is called concept-oriented programming (CoP). In CoP objects are represented and accessed indirectly by means of references. The structure of concepts describes a hierarchical space with a virtual address system. The paper describes this new approach to programming including such mechanisms as reference resolution, complex references, method interception, dual methods, life-cycle management inheritance and polymorphism.

## 1 Introduction

Let us assume that a method is applied to a local variable which stores a reference to an object. The traditional object-oriented approach to programming assumes that we do not need to know what data is actually stored inside this variable in order to have a possibility to access the referenced object. In this case we are completely unaware of the reference format and it is the task of the compiler to provide a mechanism for object access. Why this conventional approach is so convenient? The short answer is that it effectively hides almost all the numerous peculiarities of the object representation and access (ORA) mechanism and provides an illusion of instant access. Indeed, it is very convenient to simply specify a variable name followed by a field or method and then have the illusion that the result is obtained in the next moment of time.

We have so much accustomed to this approach that completely forgot that it is only an illusion and in reality each access requires significant resources in order to be executed. In other words, the software and hardware environment need to implement rather complex functionality which is responsible for object representation and access in order for the programmer to feel comfortable. If we could take a magnifying glass and look inside a reference then we would find that it is not a primitive construct at all. And if we watched closely inside the access procedure then we would learn many interesting facts. In particular, each access request results in some more or less complex sequence of actions executed at lower levels of the system organization. Interestingly, this procedure spans not only software levels but propagates further to the hardware levels. For example, in order to perform a method call in Java it is necessary to resolve this reference into a memory handle, which in turn needs to be locked and resolved into an address in virtual memory, which further has to be processed in the CPU and mapped into the physical memory possibly by loading the memory page from the swap file. And it is not the end of the story because this procedure continues at the level of micro-circuit of the memory chip and so on till the physical level of system organisation which involves electromagnetic interactions and other physical effects.

So what is the problem and why do we need to bother if the environment is already implemented in such a way that all these complications are hidden? Why do we need to know the format of references and how they are used to really access the represented objects? Shortly, the problem is that frequently in programming we would like to define our own custom reference format which is designed for representing some special class of objects. The available standard references in this case are too universal because they are implemented to serve an average object under average conditions. But if we need something special then the standard ORA mechanism may be too restrictive. Since complex software systems may involve very different types of objects with very different requirements there is a need in custom object representation and access mechanisms which are suitable for serving specific classes of objects. In other words, each concrete class of objects should be managed by the container which knows its specific features and can handle it appropriately. If we had a possibility to describe custom ORA mechanism using custom reference format and custom access procedures then we would be able to build a kind of internal virtual address space within this program. In other words, what is currently done in the hardware memory manager or operating system memory manager could be done in the program itself. These layers would be integral part of the program itself and hence they would serve the specific needs of this program. For example, we could develop a custom internal container with a special life-cycle management strategy or a container for remote objects etc.

One traditional solution of this problem consists in developing such custom containers implementing some ORA mechanism outside the program at the level of the operating system, as a middleware or library. For example, an operating system could provide additional types of object allocation mechanisms like local heap. A middleware could provide means for managing remote objects which are represented by special references and accessed using some dedicated network protocol. Or a run-time environment could provide its own container for managed objects where they are automatically garbage collected. Although these technologies can significantly help in developing complex systems we are still not satisfied with this approach. And the main reason is that all these custom containers provide *standard* ORA mechanisms which cannot be easily changed from the program. In other words, each new library or middleware for managing objects is simply *yet another* universal container. In this situation a programmer has to still rely on available standard software and cannot easily develop special containers for special classes of objects. The main problem considered in this paper is formulated in wider context. What if ORA functionality is integral part of the program itself? What if operations executed during access reflect some unique features of this and only this problem domain? What if the format of identifiers is used for the classes of objects to be developed within this program? In this case references and their associated access functions cannot be easily separated from the served objects because they are part of one program and need to be described using the same methodology and programming tools. For example, in the program we might need a huge number of very small objects frequently created and deleted with a very special allocation strategy. In addition, these objects might be distributed in a computer cluster with some special topology and hardware.

Of course, all these and other features could be implemented using traditional technologies. But the approach described in this paper is aimed at bringing them at the level of the programming language. In other words, the main goal of the proposed technology consists in providing means for developing custom ORA mechanisms at the level of an object-oriented programming language. In particular, if we have some special objects which need some special format of references and access rules then this new approach will provide means for implementing this layer using only the available programming language. Thus the hidden ORA level is made explicit and integral part of the program. References are completely legalized and get the same status of first class citizens as objects rather than permanently living in the underground. The object containers are then part of the program and the programmer is able to develop such internal containers along with other elements of this same program. The program objects are living in a virtual address space which is part of this program. The programmer is able to raise the level of indirection and the level of abstraction by making the development process more clear and efficient.

Technically the proposed approach allows the programmer to describe custom references with arbitrary structure and behaviour using the same language that is used for the rest of the program. At the same time the programmer retains the illusion of instant access, i.e., objects are used precisely as in OOP. However, such a facility has a profound effect on the whole approach to programming. The thing is that ORA functions executed behind the scenes account for a great deal and frequently most of the program complexity. Even if the program hides this level by creating the illusion of instant access these functionality is real part of any system and hence should be somehow modelled. Although the purpose of this hidden ORA level consists in serving higher (more complex) levels it can influence their behaviour. One important property of the ORA functionality is that it has a cross-cutting nature with respect to normal object methods. In particular, this means that one format of references with some access rules can be used by many different classes of objects. These functions are then triggered automatically in very different parts of the program whenever the objects served by these references are about to be accessed.

When using custom references with the associated access procedures the focus is shifted to building a container for objects *before* the objects themselves are defined. On the other hand the container has a rather complex structure which is smoothly integrated into the whole system. This means that functions of such a container are activated precisely when we need them without any explicit actions just because objects involved into the interaction have some position within this structure. The program in this case is analogous to a hierarchical address system like postal addresses. Each object within this system has a unique address which is a convention defined by the programmer rather than an address in memory or another kind of native identifier. Accordingly, when such an object is about to be accessed it is necessary to resolve this address. The same happens in the real world when we want to perform some actions. For example, in order to open a new bank account we write an application and then send it to the bank by specifying its address. In some time we have our account opened and have the illusion of instant action. However, it is not so and this illusion appears because the transport level is separated from the main business logic. In fact, the application was processed at the local post office then passed through some intermediate post offices and finally was processed at the bank itself before the business method for opening new bank accounts could be started. In the bank itself the application could pass through several departments responsible for different operations. An interesting observation here is that the complexity of intermediate processing may well be higher than the final operation. Another property of such a system is that all the levels have to be integrated and all the components constitute one whole. In the proposed approach such integrity and common rules are enforced at the programming language level by introducing new language constructs and other language means. Thus the programmer gets significant support from the programming language when describing such object containers with virtual address system.

The rest of this paper is organized as follows. In Section 2 we introduce main notions underlying the proposed approach by describing properties and responsibilities of objects and references (Section 2.1), indirect access via reference resolution (Section 2.2) and indirect access in a hierarchical address space (Section 2.3). In Section 3 we define a new programming construct, called concept, which is the basis for the whole approach. Concepts generalize classes and allow the programmer to describe a hierarchical address space with indirect access to objects. Just as classes concepts live in a hierarchy which is specified using inclusion relation described in Section 4. Inclusion relation between concepts generalizes class inheritance and is used to model hierarchical space of objects and references. Operations with references are described in Section 5 while the mechanism of life-cycle management is described in Section 6. Section 7 is devoted to the analysis of how inheritance and polymorphism are working in the new approach. Dual methods are described in Section 8 while Section 9 is aimed at analysis of related work. In Section 10 we describe how the proposed mechanisms change the whole approach to programming and system design. Section 11 concludes with a summary of the paper.

Throughout this paper we use the following conventions. Basic or simple elements in a hierarchy are positioned higher than more complex elements like class extensions which are positioned lower. Objects and classes of objects are shown as (white) rectangles while references and classes of references are shown as (gray) rounded rectangles.

## 2 Object Representation and Access

### 2.1 Objects and References

References are used as a means to represent entities. A reference acts as a link to an entity and in this sense it is able to identify entities. Here it is implicitly supposed that the only way to access an entity consists in getting and using its reference. Thus all things can be divided into two kinds which are referred to as *entities* and *identities*. Two realms, the realm of entities and the realm of identities, coexist. However, how they are connected is actually a magic. The interactions between them propagate infinitely deeply into the physical structure of the matter down to the lowest levels of its organization. We overcome this problem by apply the standard trick: we stop at some level and declare it a basic or axiomatic one without any further explanations. For example, we might choose quantum interactions as such a level or, more realistic for computer science, network protocol level or memory operations level. Once such a basic level is fixed it is possible to build a theory of interactions between the two worlds of entities and identities at higher levels. In this case higher level interactions are reduced to the basic interactions however the separation between the two realms remains and plays an important role for the whole approach. Further in the paper we will use terms *object* and *reference* which are generally accepted in computer programming instead entity and identity, which are used more in data modelling.

Since references are used to distinguish objects they need to contain unique information. This information stored in a reference can be viewed as a coordinate of the represented object in some space. Thus we come to an important conclusion that objects and references do not exist in vacuum.

Rather they need an additional element which is called space. The space can be thought of as an environment or container which is associated with a set of objects and their references. It is important that the existence of space is a necessary requirement and considering objects or references out of the space context simply does not make sense. Space in our approach is not simply a passive scope but rather an active environment which can influence what happens inside and change properties and behaviour of internal objects. In particular, the space is responsible for inter-object communications and for managing their life-cycle. The space is precisely the medium that allows interactions to propagate. Since objects have different coordinates they need some intermediate environment which can transfer interactions between them. Such movements cannot be executed instantaneously and this is why any object access is indirect and takes some time to execute. Information in the reference can be interpreted as an encoded path to the represented object which needs to be passed before it is possible to affect the object.

References and objects are two types of elements constituting any program which live each in their own realms. However, both references and objects need to contain information in a structured form and this does not depend on their role in the program. In traditional programming references are primitive elements and even if they have some internal non-primitive structure the programmer does not see that and has no means to model them. In contrast, in the concept-oriented programming references have the same rights as objects and hence it is necessary to have means for describing their structure just as we do it for objects. For example, one reference might contain information indented to identify persons like name, passport number and birth date. Other references might be designed to represent companies including such information as name and registration number. Structure of information may have very different forms but in computer programming the wide-spread approach consists in using named fields which are used to break all the information onto smaller pieces and methods for describing behaviour associated with this information. A construct used to describe state and behaviour using fields and methods is referred to as a *class*, which is a basic language element in object-oriented programming. The concept-oriented programming does not add anything to this approach and still uses classes to describe state and behaviour. What is new is that now classes are used to describe references in addition to objects. Thus the concept-oriented programming uses two types of classes: an *object class* and a *reference class* (Fig. 1). Object classes are used as usual to describe the structure and behaviour of objects while references classes are used to describe that of references. Accordingly, instances of reference classes are referred to as *references* while instances of objects classes are referred to as *objects*.

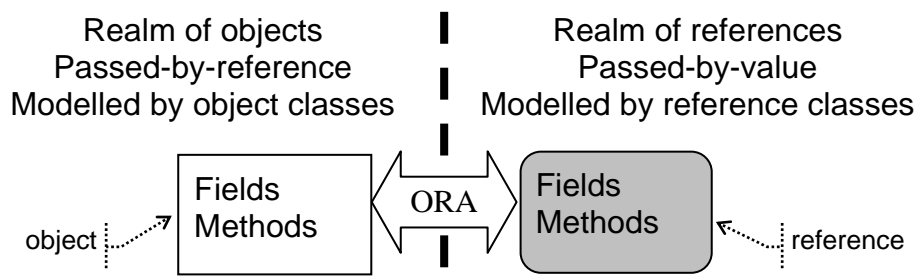

Figure 1. The realm of references and the realm of objects modelled by reference classes and object classes.

Reference classes do not differ from object classes except that they are marked as a reference class. All other facilities of conventional (object) classes can be used within reference classes. In particular, they have fields, methods, constructors, destructors etc. For example, a reference class for identifying bank accounts could be defined as follows:

```
reference AccountReference {
  String accountNumber;
  Date openingDate;
  String getAccountNumber() { return accountNumber; }
  double getBalance() { ... }
  ...
}
```

Here keyword 'reference' is used to declare this class as a reference class while for declaring object classes we will use keyword 'class'. This class stores an account number as a text string in the first field and an opening date in the second field. It also provides some methods implementing functionality associated with these references.

The main difference of reference classes from object classes is that their instances (references) are passed-by-value. Reference is a means of *representing* objects and as such it does not allow having its own reference. In other words, a reference represents itself by its own contents and the only way to pass a reference consists in copying information that is stored in it. One consequence of this property is that references are used as an information transfer mechanism. If it is necessary to transfer information from one location to another then the only way consists in using some reference. In contrast, objects are passed-by-reference and never change their position in space. If an object has been created with some reference as its representative then the position of the object cannot be changed because this will entail the necessity to update all the references stored in other objects. Thus references can be though of as travelling elements while objects always retain their position in space. In particular, the assignment operator is implemented by copying the reference stored in the variable while the represented object is not involved into this process. For example, if it is necessary to copy an account from one variable to another variable then we write it as follows:

```
AccountReference account2 = account1;
```

This statement means that information stored within the first variable is simply copied to the second variable. However, in CoP this information may have any structure which is defined by the corresponding reference class. If these two variables are declared as having class `AccountReference` then two fields have to be copied from the first location to the second one.

In the concept-oriented program, references are passed-by-value and provide a mechanism for information transfer and object access. It is important to understand that anything that happens in the program relates first of all to references rather than objects. Objects are protected from any direct action by their references. Such a situation takes place in almost any existing approach to programming and what is new in the concept-oriented programming is that this barrier implemented by references is made explicit and legal. In particular, it is important to understand that any operation applied to an object starts at the level of its reference. In other words, any access to an object is intercepted by the reference that represents it. For example, assume that an account is represented by its reference stored in some variable. If we apply some method to this variable then the first method to execute will be that of the reference class:

```
String accountNumber = account1.getAccountNumber();
```

Here we applied this method to a variable but the reference will intercept this call and return the account number stored in this reference while the represented object is not involved into this operation.

Reference acts as a passed-by-value proxy for the represented object. The interception of all access requests by the reference does make sense because it is the only element in the program that actually knows how to reach the represented object by interpreting information stored in its fields. Informally, any access is interpreted by the reference as a request to do something with the represented object. And although references create the illusion of accessing objects directly they always intercept any access. In the existing approaches this indirection is hidden while in CoP it is made part of the main logic of the program. Any access request in CoP does not directly reach the target object but is intercepted by its reference and then delegated further to the object. For example, if we need to get the current balance of the account then we apply the corresponding method to its reference:

```
double balance = account.getBalance();
```

However, this reference stores arbitrary data such account number and the account object cannot be accessed directly because it can reside anywhere in the world. The reference does not store a memory address or any other kind of native reference so such a representation is indirect. To access an indirectly represented object it is necessary to perform some additional actions by finding the actual location of the object (Fig. 2). For example, the account number could be loaded from persistent storage in memory and then its memory location is used as a direct reference that can be used for executing the target method and getting the account balance. Because of the necessity in such intermediate actions this access is called indirect. Thus CoP can be viewed as a way of describing actions that happen behind the scenes during object access. In OOP these operations exist but cannot be described in the program. In CoP they are made integral part of the program.

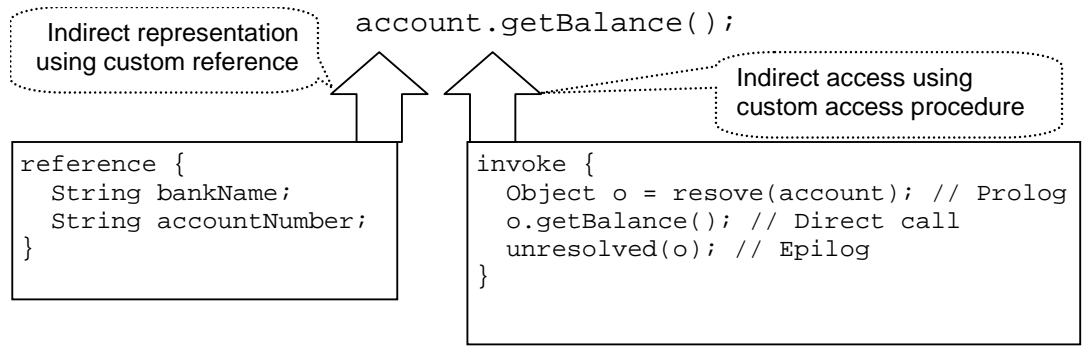

Figure 2. In CoP objects are represented and accessed indirectly.

### 2.2 Indirect Access using Reference Substitution and Resolution

Let us consider a very simple example where we have a computer with some IP address and want to provide some services to our clients. The IP address is a reference and hence we give it to our client in order to access our computer from outside. So what is wrong with this approach? Obviously the problem is that the IP address may change in future (for example, if we change our ISP) and then we need to notify all our clients. In order to overcome this difficulty we can introduce an additional level of indirection where computers are identified by names which are more stable. However, names cannot be used to access computers and hence we need to store somewhere a mapping from computer names to their corresponding IP addresses. Thus we still use IP address for direct access however computers are identified by names which are stored by all the clients. Notice that IP address appear only in the name resolution service and clients are completely unaware of them — they have an illusion that the name provides them access to the service computer. In this example we used the mechanism of reference substitution. The idea is that references in new custom format are used to represent objects instead of old references. Each such new reference (computer name) replaces one old reference (IP address) and this mapping is maintained by some well known service. When it is necessary to access an object the source object asks the resolution service to resolve this object reference and the result is then used for direct access (Fig. 3).

Let us now continue this example and assume that we need to move our service to another computer in the network along with the old IP address. This means that both the name of the computer and the IP address do not change however this old IP address is used now by another computer. So how the clients find our service if it already resides on another computer? The solution is that the IP address is now mapped to another computer physical address, called MAC address. An important consequence of this example is that the IP address is not a final reference (although from the point of view of names it is viewed so). When we get an IP address we need to resolve it to a physical MAC address in one or another point of our access procedure. Accordingly, IP address substitutes for MAC address and is resolved into it whenever a computer is about to be accessed. It is also important to notice that this process can be continued infinitely through different levels of the system organization. It does not stop at software level and continues further to hardware levels and even further to physical organization levels. However, in system design we normally choose some functionality as a basic level. After that this basic functionality is used to describe more complex and more indirect levels. For example, contemporary processors provide a rather complex memory addressing system which is based on the hardware memory manager. This system is controlled by the operating system which is able to access any address without any restrictions. The operating system creates a level of indirection over the hardware physical addressing by proving some other format of reference to application program. It can be a kind of memory handle or a segment-offset pair but in any case this application oriented memory references cannot be directly used to access the memory. Instead, they must be resolved into the corresponding physical address. Further, the application program or a run-time environment may organize its own levels of indirection. For example, Java provides its own format of references which substitute memory handles.

Although the described example of IP address substitution and name resolution may seem trivial we describe it because of its importance for the whole concept-oriented approach. This example shows that there is some functionality which must be executed even if it has not been requested by the programmer. For large systems these functions can reside on many levels of system organization and can account for a great deal of the system complexity. And the main goal of CoP consists in providing means for describing these hidden functions, particularly, the functions of reference resolution and substitution.

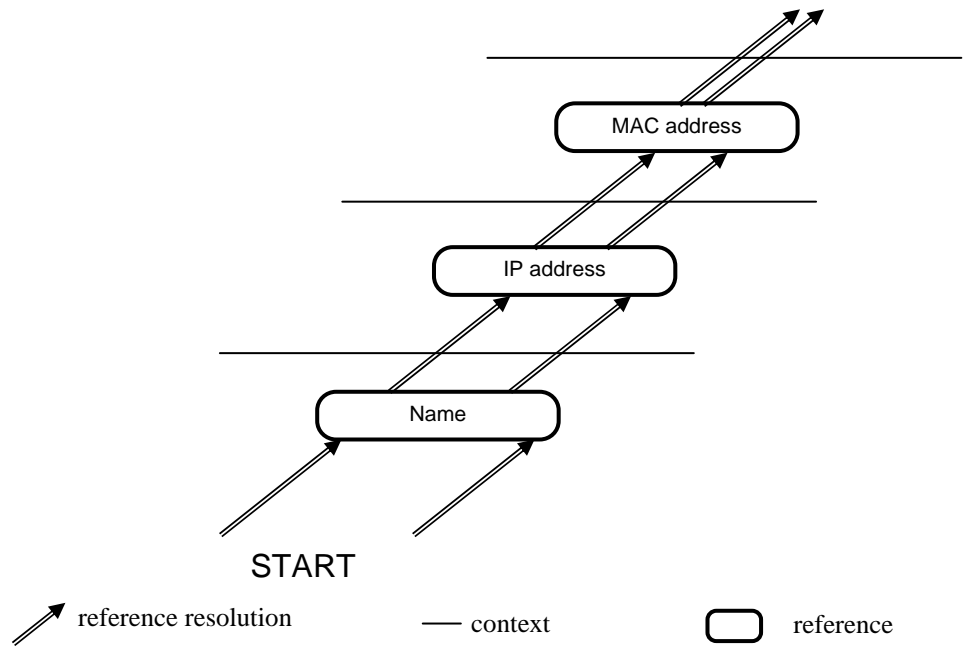

Figure 3. A sequence of resolution of computer names.

Let us consider another example where a variable in a Java program contains a reference to an account object and this variable is used to get the current balance of this account:

```
double balance = account.getBalance();
```

From the point of view of the programmer this statement results in an execution of the specified method without any intermediate operations. In other words, the next line of code will be the first statement of the invoked method defined in the object class. However, if we look under the hood, then we will see that the method call is not actually a primitive operation (Fig. 4). The thing is that this reference is a Java reference and hence it contains an arbitrary identifier of the object. The value stored in the reference cannot be used for object access because it is simply a convention used within the Java virtual machine. In order to find the object it is necessary to convert this reference into a memory handle. The memory handle is allocated by and used at the level of the operating system for representing objects in memory (more precisely, objects in the global heap). However, memory handles are also conventions and they cannot be used directly to access an object in physical memory. Instead, it is necessary to lock it and get an address in memory where this object resides. This address can be then used by the processor to access the object in memory. Thus any object access implicitly results in a rather complex sequence of intermediate actions which are not visible by the programmer. Notice that one and the same access procedure is execute for each object access.

Interestingly, getting an address in memory is not the end of the story. The processor takes this address and has to map it to physical memory by loading the corresponding memory page from its swap file if necessary. The memory itself executes some operations at the level of its microcircuits and so on down to the physical levels of the system structure. In this case each level can be chosen as the basic one while higher levels remain visible. In this way we can vary the level of abstraction by hiding unnecessary details. However, it is necessary to understand that there exist always some procedures at lower levels and the only question is whether they are visible or not.

In object-oriented programming it is assumed that reference format and access procedures are managed by some routines that are not directly controlled by the programmer. For example, memory handles are allocated by the operating system and Java references are provided by the Java virtual machine. In such an approach the programmer is unaware of how objects are represented and what intermediate actions are performed behind the scenes after a method is called and before its first statement starts executing. Such a situation is a consequence of the object-oriented principle which postulates that a system is a set of objects, i.e., all the system functionality is encapsulated in objects. According to this principle, in order to describe a system it is enough to define its classes of objects. In particular, any function that is being executed within the program is the result of some *explicit* method invocation written by the programmer somewhere in the source code. And any object that appears in the program is the result of an *explicit* instantiation. There are no other objects and method calls except those specified by the programmer. Under such an assumption any reference looks like a primitive element without any internal structure and any method call looks like an instant action. In other words, object-oriented approach creates the impression of instantaneous method invocations without any intermediate operations.

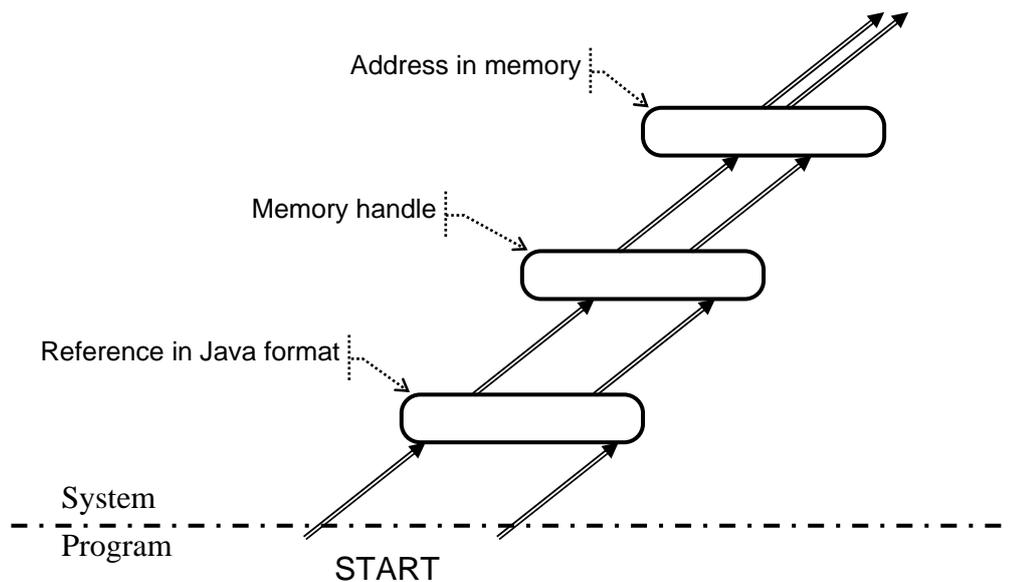

Figure 4. A sequence of resolution of Java reference.

Obviously, there is nothing bad in hiding unnecessary details and providing the illusion of direct access to objects. Moreover, such an approach where the programmer does not have to care about object representation and access was a significant advantage over the previous technologies. The main problem with this approach is the absence of any means to modify the format of representation and access procedures. Hiding all the details and peculiarities in OOP means completely removing this level from the program and keeping only one or a few standard ORA mechanisms. Indeed, we do not need and do not want to know at all how the access procedure works when calling a method. But what we really would like to have is to be able to intervene somehow into this process. For example, what if I want to print a message whenever an account is about to be accessed? Probably the only way consists in breaking into the Java virtual machine and inserting one line of code with this output message into the resolution procedure. Since it is not possible we have to insert this line of code before each access to any account object in the program. What if we need to define our own format of references which are used to represent some objects? In this case again OOP provides very limited support. In general the requirements to the object representation and access mechanism may be quite complex and very specific. For example, a system might need to carry out special security checks whenever its objects are accessed. Or, a program uses a huge number of small objects with very short life time which in addition depends on the state of the program. In all these and many other cases the standard mechanism of representation and access could be too restrictive because it cannot be modified and adapted to the concrete needs of this program.

Let us now consider how references are used in the concept-oriented programming. Shortly, the main difference is that the programmer is able to introduce new levels of indirection within this same program by defining custom references. If in OOP different levels are positioned outside the program then in CoP it is possible to build new levels of indirection within this program as its integral part. A new reference class will produce references which substitute for native references. It is possible to build a hierarchy where each new reference will substitute for an existing reference up to the system reference. Accordingly, when an object represented by such a custom reference is about to be accessed this reference is resolved into the substituted reference in a nested manger up to the system reference.

For example, let us assume that we want to use custom references for representing bank accounts as was described in the previous section. The reference class contains a field with the account number and hence such a reference has to be resolved before the account can be used (Fig. 5). However, accounts could be persistently stored in a database where each object has some unique identifier like primary key. Each account number is then resolved into its primary key. This primary key is also a convention and cannot be used for object access. So we have to proceed further by loading the account object from the database into memory and restoring its system reference. This reference can be then used to access this object. Thus the whole procedure is still wrapped into a nested sequence of reference resolutions but the difference is that now it is part of the program because the custom references are defined within this same program using a reference class.

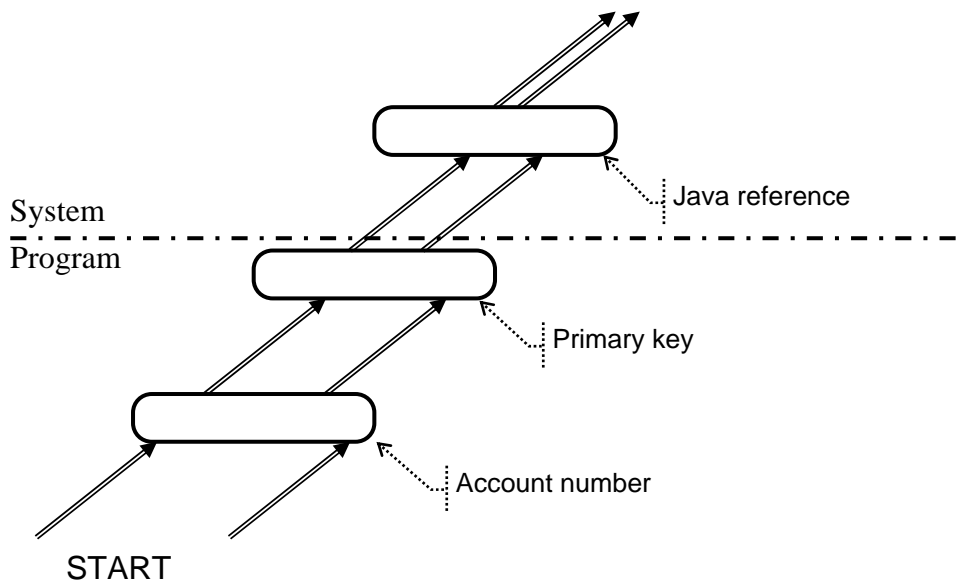

Figure 5. A sequence of resolution of an account object given its custom reference.

### 2.3 Hierarchical Access using Reference Concatenation and Segmentation

In the previous section we described the mechanism of reference substitution and resolution (RSR) which allows us to interpret some data as an indirect representation of an object. When an object is being accessed its reference is resolved into a simpler reference which in turn is resolved in another reference and so on up to the level that provides basic references with direct access. It is typical for the concept-oriented approach that there exists also a dual side for a property or mechanism. In this case the dual mechanism is referred to as reference concatenation and segmentation (RCS). This mechanism describes the hierarchical structure of the space where objects reside. Indeed, the RSR mechanism assumes that an object is represented by some reference. However, this reference makes sense only in some scope or context and then the issue is that this context should have its own representation using some reference. Thus any reference must have a higher level reference or higher segment in order to be interpretable. For example (Fig. 6), in order to interpret a street name it is necessary to know the city name which is its context. And in order to meaningfully interpret this city name it is necessary to know the country and so on.

On the other hand, a reference needs not to represent the final object. Instead, it can represent an intermediate object in the context in which the final or the next object exists. Thus a reference may have a lower level reference or a lower segment which represents the next object in the sequence of access. For example, if a reference represents a street then we can use it to access this street object. However, the street reference may have the next segment representing a house on this street.

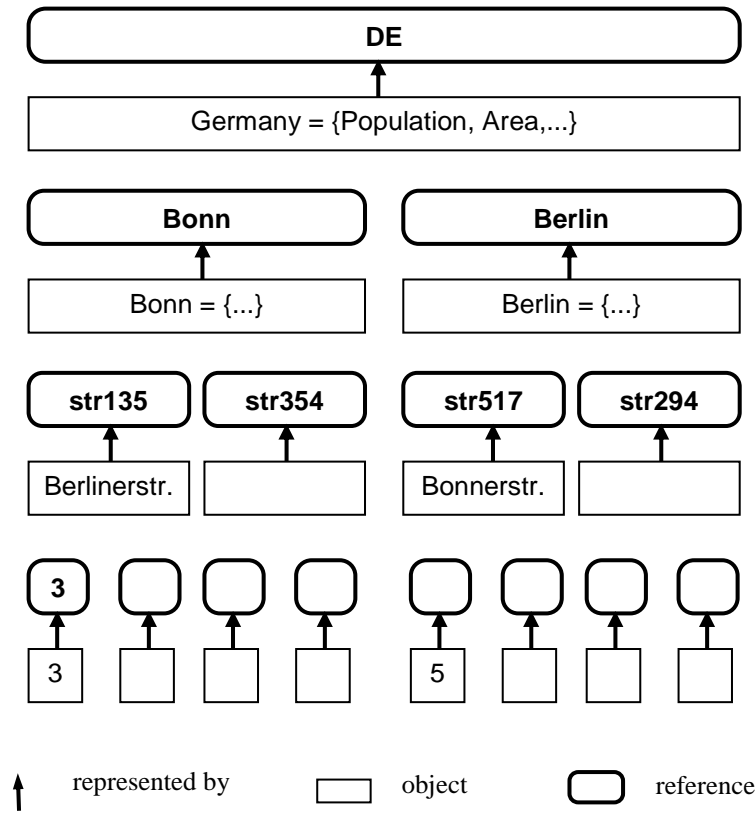

Figure 6. An example of a hierarchical structure of objects and references.

According to this reasoning any object can be viewed as located in a nested (hierarchical) system of spaces like Russian doll. Both the objects and the intermediate spaces are represented by some reference, which distinguishes them from other objects or spaces. However, the main point of the RCS mechanism is that any such reference needs some higher segment representing its parent space and may have a lower segment representing a child object or space. Notice that here we do not distinguish spaces and objects because both of them have some reference in the context of the parent space. The only difference is that an object reference does not have the next segment concatenated to it.

We say that the parent reference and the child reference are concatenated to this object reference from two sides by pointing to the parent context and the internal object, respectively. A sequence of references where each previous segment represents the parent object for this reference is referred to as a *complex reference*. An example of a complex reference is a postal address where each segment represents an object in the context of the object represented by the previous segment. Another example of a complex reference is a computer name where segments are separated by dots. For example, www.conceptoreinted.com consists of three segments where com is a high segment and www is a low segment. Notice that all the segments are indirect identifiers which need to be resolved using RSR mechanism described in the previous section. In particular, in order to access the computer represented by its name www.conceptoriented.com it is necessary to resolve all its three segments. The natural sequence of access on complex references starts from the high segment and ends with the low segment. Higher segments represent external spaces which must exist *before* internal elements can be defined. Therefore higher segments and the corresponding elements can be viewed as base units while internal elements are more complex elements which cannot exist without the base elements. And here we see that the natural sequence of access on a complex reference starts from the base element (high segment) and ends with the most complex element (low segment). In contrast, when resolving references we start from the most complex reference and then proceed to simpler references till the base reference. So these two mechanisms, RSR and RCS, have an opposite direction in their sequences of access.

## 3 Concept Definition

### 3.1 Reference Class and Object Class

In the previous sections we postulated that any program consists of two types of elements: references and objects. Their functions and behaviour can be described by means of reference classes and object classes, respectively. However, the main problem is that using *separately* reference classes and object classes hardly makes sense because they are not unrelated elements but rather always co-exist and co-operate. In other words, they are dual but indivisible elements and hence we need some methods for handling their indivisible unity.

In order to solve this problem we propose to use a new programming construct, called *concept*, which is defined as a pair consisting of one reference class and one object class. Thus reference classes and object classes are not used as individual constructs anymore. Instead of them, the programmer has to define concepts. Once a concept has been defined it can be used to declare a type of variables, parameters, fields, return values and other elements of the program where in OOP classes have always been used. Reference class and object class loose their individuality and can be used only via concepts. For example, if we want to define a normal class then it is necessary to define a concept with the empty reference class. If we need to define a reference class then the object class of the concept has to be empty. On the other hand, a concept can be thought of as a normal class with an additional reference class attached to it and in this sense concept is a generalization of class.

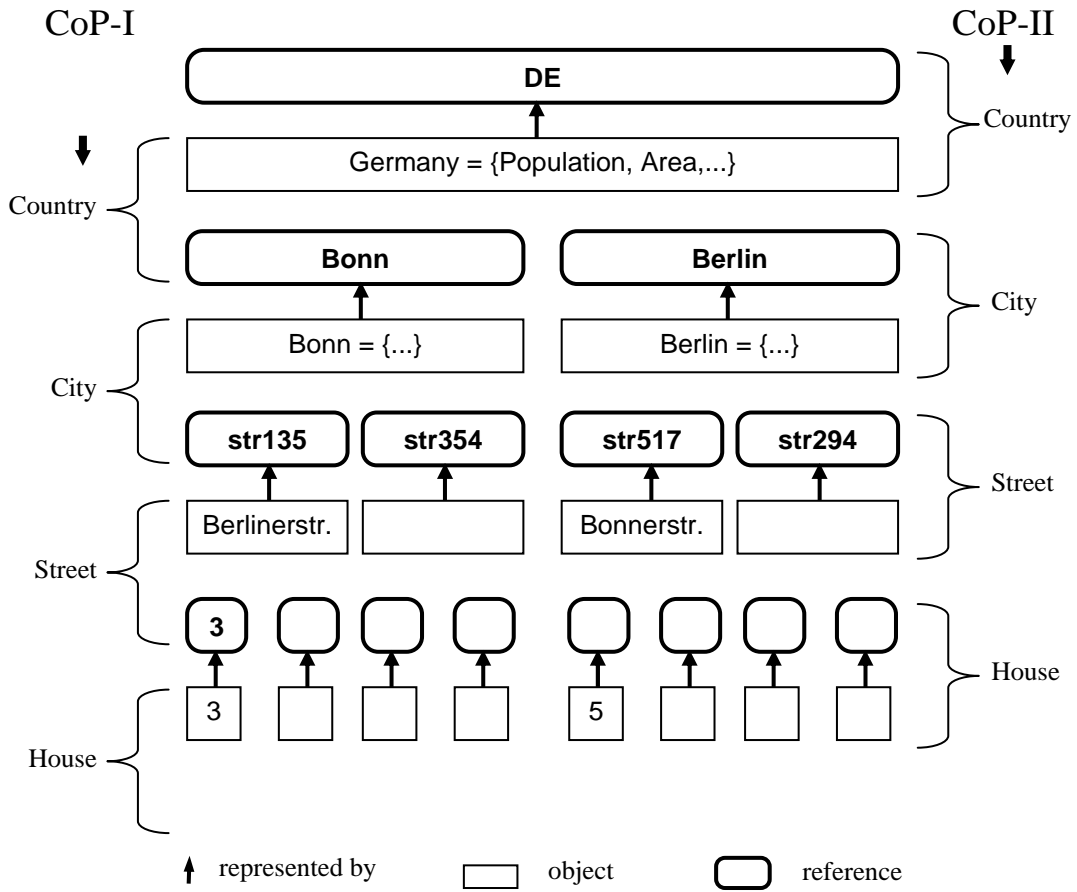

Figure 7. Two approaches to concept composition: CoP-I (left) and CoP-II (right).

Defining concept is a pair of two classes is not enough — it is necessary to define their mutual roles. In other words, it is necessary to define how instances of the reference class relate to instances of the object class. If we provide such roles then concepts can be used to describe the space of objects and references. If we want a concept to describe neighbour elements of the reference-object hierarchy located next to each other then there are only two major alternatives denoted as CoP-I and Cop-II (Fig. 7). These alternatives depend on how we assign two classes of one concept to objects and references of the space.

The first alternative assumes that if a reference has a class of one concept then the object it represents has a class of the child concept (Fig. 7, left). And vice versa, if an object is described by a concept then its reference is described by its parent concept. Thus objects are represented by parent concept references and references represent child concept objects. For example, city objects are described by the object class of the City concept while city references are described by the reference class of the Country concept which is a parent of the City concept. In other words, city objects are described by one concept while city references are described in the parent concept. This approach was studied in the previous papers on the concept-oriented programming [Sav05a] and is not described in this paper.

The second alternative assumes that a reference and an object represented by this reference are described within the same concept (Fig. 7 right). This means that when defining a new concept we describe a class of objects and simultaneously a class of references for representing these objects. Thus objects are represented by references of this same concept and references represent objects of this same concept. For example, the City concept will have an object class that describes city objects while its reference class is used to identify these cities. This approach is described further in this paper.

Listing 1 provides an example of a concept definition. It starts from the keyword 'concept' followed by the concept name. The reference class starts from the keyword 'reference' and includes a filed intended to identify objects of this concept. The object class starts from the keyword 'class' and includes one field which stores the state of the object. When an instance of this concept is created then normally a pair of elements is created: a new reference and a new object. The difference is that the reference is stored by value directly in the variable while the object is created in some storage (not necessarily in memory). Thus any variable declared as having this concept as its type needs to allocate a space on the stack which is enough to store a reference of this concept. In contrast, in OOP all variables allocate the same space on the stack because there is only one type of reference having the same standard length. For example, variable `account` (line 11) stores an account number rather than a native reference to the account object. If we need to pass this variable as a parameter to a method then (line 12) this reference is copied, i.e., the account number is passed-by-value to the method while native reference is not involved and in fact may not even exist at this time.

**Listing 1. An example of concept definition.**

```
101  concept Account
102    reference { // Reference class
103      String accountNumber; // Object identifier
104      ...
105    }
106    class { // Object class
107      double balance = 0;
108      ...
109    }
110
111  Account account = new Account();
112  doSomething(account);
```

Although references are supposed to represent objects it is not strictly necessary. It is important only that instances of the reference class are passed by value. For example, some objects have to be passed by value although they do not represent anything but simply contain some information. Or we might want to include some additional information in a reference for convenience.

The first question that arises here is how to distinguish between reference class members and object class members. Indeed, we define them as two separate classes within one concept but use them as one element of the program because there is only one name for the both. In particular, the reference class and the object class of one concept may have the same members, called *dual members*, and they need to be somehow distinguished. When a member of a concept is accessed, it is not clear if it belongs to the reference class or to the object class. In order to resolve this ambiguity we use the following principle: *reference intercepts any access to the object*. This means that if we read/write a field then the reference tries to intercept this operation and if this field exists in the reference class then its value is read/written. If a method is applied to a variable then the reference tries to intercept this call and if this method is defined in the reference class then it is executed. This principle is quite natural because references serve as object representatives and hence all access requests need to pass through them. Another simple reason for this rule is that objects in the general case cannot be accessed without reference intervention because only reference knows how to access them. References store not only the object address but also the logic of its interpretation (resolution).

According to the interception principle references protect objects from direct access. Then the question is how an object can be accessed if all requests are intercepted by its reference? Here we assume that once a reference method has intercepted a method call, all other accesses from within its context are directed to this object. In other words, it is assumed that references are able to perform direct access to their objects. If a reference represents a border then this border intercepts all attempts to get into its scope. Reference methods are interpreted as tunnels or passes through the border with different processing algorithms and once the request is within this tunnel (within some reference method) it is possible to go further inside in the direction of the target object (Fig. 8).

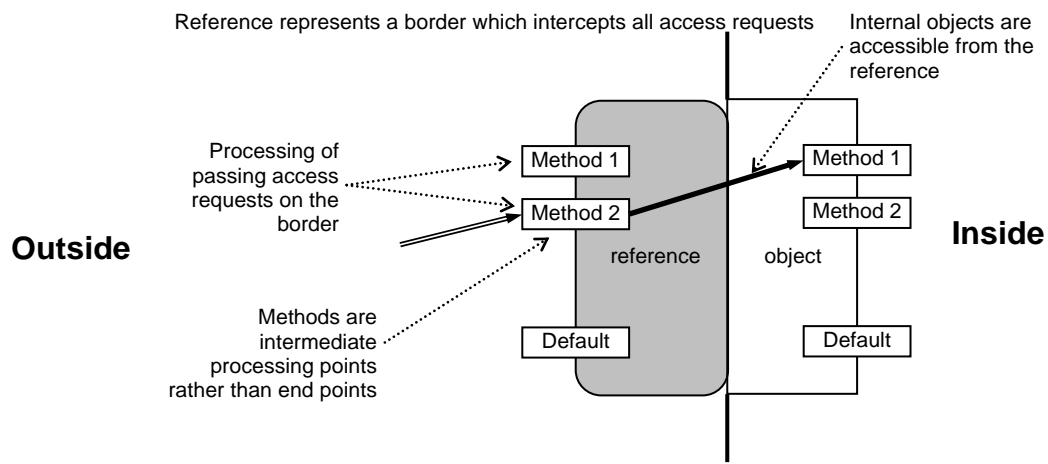

Figure 8. Reference intercepts all accesses to the represented object.

A reference can access its object members by specifying their names. Let us now assume that an object method is being executed and we need to access its reference. Here again we need to have some disambiguation technique because any access within a concept can be interpreted either as the reference member or the object method. We will assume that reference members are distinguished from object methods by using the keyword 'this'. In other words, 'this' keyword designates the current reference and hence any method or field applied to it refers to that of the reference.

Let us now consider a simple example illustrating these principles shown in Listing 2. Here both the reference class and the object class have a field storing an account balance (lines 4 and 10). Assume that a new account object gets zero balance which is also stored in its reference (as a cached copy). However, shortly after its creation the account gets 5 EUR bonus. Thus all the references still store the initial (old) zero balance while the object stores the new (true) balance. If we read the balance field of the account reference then we get the old zero balance because the object field is effectively hidden (line 16). Since references are passed by value they all store their own field values. In order to get the value stored in the reference it is possible to write a special method of the reference class (line 6). However, in order to indicate that we need a field in the reference rather than in the object this method uses the keyword 'this'. If we apply this method to the account object (line 17) then it returns the values stored in its reference.

In order to get the current balance stored in the object we can write a special method of the reference (line 7). It actually does nothing and simply delegates this request to the corresponding method of the object (keyword 'this' is not used). The method of the object class returns the value of its field (line 12) and here notice again that we do not use the keyword 'this' by indicating that the member of the object class is intended. If we remove the method `getBalance` of the reference class then the equivalent behaviour will be produced by default. In other words, if the method applied to a reference is not defined in the reference class then the default behaviour is to call the same method of the object class.

**Listing 2. Distinguishing between object class and reference class members.**

```
201  concept Account
202    reference { // Reference class
203       String accountNumber; // Object identifier
204       double balance;
205
206       double getLastBalance() { this.balance; }
207       double getBalance() { return getBalance(); }
208    }
209    class { // Object class
210       double balance = 0;
211
212       double getBalance() { return balance; }
213    }
214
215  Account account = getAccount();
216  double balance = account.balance; // = 0.00 EUR
217  double last = account.getLastBalance(); // = 0.00 EUR
218  double current = account.getBalance(); // = 5.00 EUR
```

A consequence of the interception principle is that the computations in the program are actually calls of reference methods. Even if a reference class does not define some method we can always assume that it exists and performs some default action. There is no way to bypass a reference if it wants to control access to the represented object. So reference implements the logic of a border or intermediate environment. Notice that the programmer has still the illusion of working with the object directly so the requirement of transparency is satisfied. In OOP and other existing approaches to programming this intermediate behaviour is not controlled by the programmer while in CoP it is completely legal and plays an important role in system design.

### 3.2 Reference Resolution

In the previous section it was assumed that reference methods can call object methods using their name (without the keyword 'this'). For example, the reference method `getBalance` could call the dual method of the object class as follows:

```
reference {
  ...
  double getBalance() { // Reference method
    return getBalance(); // Object method call
  }
  ...
}
```

However, in order to call an object method it is necessary to know this object direct (native) reference. In this example it is assumed that the variable contains a reference in the custom format as an account number (instead of a native reference in OOP). Then the reference method can be called in the context of the reference which is stored on the stack. However, strictly speaking, it is not possible to call the object method because we simply do not know this object native reference. Indeed, the only thing we know is the account number and that is all. The account object itself can be anywhere in the world and not necessarily in memory. Thus an object method cannot be called from the custom reference without an additional support for finding this object native reference. In this section we describe how this support is organized.

Custom references are used for creating a level of indirection and unbinding object identifiers from the hardware and software reality. However, any interaction with the object is possible only using this reality, i.e., we still need to restore somehow the object native reference in order to access it. Thus any reference substitutes for a native reference which needs to be resolved each time this object is going to be accessed.

The reference resolution functionality is encapsulated in a special method of the reference class, called a *continuation method*. The main goal of this method consists in interpreting the fields of this reference and translating them into the native reference which directly represents the object. Thus the continuation method reconstructs the missing link between an object method invocation and the possibility to execute it.

Let us consider an example shown in Listing 3. Here it is assumed that account objects identified by their numbers reside in some persistent storage. In order to interact with such an object it is necessary to load its state in memory where it is represented by a native reference having class `Object` (line 8). After the interaction is finished the state of the object can be saved back to the storage (line 10). This method is written by the programmer but is supposed to be called automatically whenever an object of this concept needs to be accessed. Actually, it does not matter how the native reference is reconstructed. We can use look up table, or some index or any other resolution algorithm that is suitable for finding the mapping between custom references and native references. For the compiler it is only important that such a method is available and that it generates somehow the native reference.

Returning resolved references in the source context is rather bad practice (it is insecure and error prone). On the other hand, native references resolved by the continuation methods are not used explicitly by the programmer. Instead, they are used automatically by the compiler when calling object methods. Therefore the continuation method does not return the resolved value but marks the place where it is available for further use. As such a mark the continuation method simply makes a recursive call which is applied to the resolved reference (line 9). In this example the resolved reference is of native type and hence the compiler will call the object method. In other cases the resolved reference may have some other type and then the corresponding continuation method will be called. Thus line 9 is precisely where the target object method should be called.

**Listing 3. Reference resolution using continuation method.**

```
301  concept Account
302    reference { // Reference class
303      String accountNumber; // Object identifier
304      double balance;
305
306      void continue() {
307        print("> Account: Start resolving account " + accountNumber);
308        Object o = load(this.accountNumber);
309        o.continue();
310        store(this.accountNumber, o);
311        print("< Account: End resolving account " + accountNumber);
312      }
313
314      double getLastBalance() { this.balance; }
315      double getBalance() {
316        print("=== Account: Reference method is called");
317        return getBalance();
318      }
319    }
320    class { // Object class
321      double balance = 0;
322
323      double getBalance() {
324        print("--- Account: Object method is called");
325        return balance;
326      }
327    }
328
329  Account account = getAccount();
330  double current = account.getBalance(); // = 5.00 EUR
331
332  $ === Account: Reference method is called
333  $ > Account: Start resolving account 123456789
334  $ --- Account: Object method is called
335  $ < Account: End resolving account 123456789
```

Given the continuation method a concept can be used to instantiate and use its objects because now references are connected with the objects they represent. For example, if we create a new account (line 29) and then call some its method (line 30) then this concept will produce the output shown in lines 32-35. Here we see that the reference method is called without any indirection while the object method is wrapped into the continuation method. This wrapping has been performed automatically and the programmer still uses the objects as if they were directly accessible OOP objects (the principle of transparency). On the other hand, the programmer has full control over the intermediate behaviour which is modularized in the continuation method.

The continuation method provides means for reaching the realm of objects from the realm of references (Fig. 9). The point in the program where the control is passed from a reference to an object is referred to as a *turn point*. Theoretically, this process is some magic because it must proceed infinitely deeply into the physical reality. In practice it is solved by resolving this reference to a simpler reference which is declared to provide direct access to the represented object.

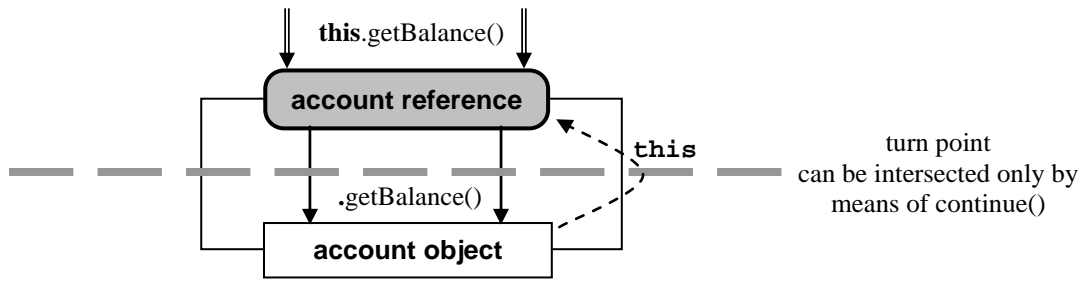

Figure 9. Turn point as a pass into the realm of objects implemented via the continuation method.

It should be noted that the continuation method is provided by the programmer and can use any resolution strategy or include any other necessary code. It is important only that the compiler will follow the concrete sequence of steps when an object is going to be accessed. In particular, the continuation method may include more complex logic then simply object resolution. Its general purpose consists in providing a mechanism for border intersection and code that will trigger automatically whenever a process wants to intersect this border. This is precisely the code that is hidden in the conventional object-oriented programming.

# 4 Concept Inclusion

## 4.1 Complex References

Concepts are not defined in isolation just like objects cannot live in vacuum without an external space or environment. In order to model the hierarchical nature of the space where objects live each concept needs a *parent* or *super-concept* to be specified, which is also called a *base concept* by analogy with base classes in OOP. We say that the concept is included into its parent concept because the space it describes is included into the space described by the parent concept. The inclusion relation between concepts defines the sub-concept/super-concept hierarchy. If concept $S$ is included in concept $C$ then we write it as follows: $S < C$ ($S$ is less than $C$). Sub-concept is more specific that its super-concept while super-concept is more general than its sub-concepts.

In the example code we will assume that parent concepts are declared by means of the keyword 'in'. For example, if concept C is included in concept B, which in turn is included in A then it is written as follows:

```
concept A
  reference { ... }
  class { ... }
concept B in A
  reference { ... }
  class { ... }
concept C in B
  reference { ... }
  class { ... }
```

The inclusion relation between concepts is analogous to the inheritance relation between classes. In particular, if concepts do not use reference classes then the inclusion relation is reduced to the class inheritance. Let us now consider what happens if an instance of concept `C` is created where concept `C` has a parent concept. In this case the new reference will inherit all parent concept references precisely as it is done in OOP. Such a reference is referred to as a *complex reference* while individual references it consists of are called *segments*. Thus a complex reference is a sequence of reference segments starting from the first parent (high segment) and ending with the reference (low segment) belonging to the declared concept. In our example the new reference will contain three segments `A` (high segment), `B` (middle segment) and `C` (low segment). All these segments are concatenated and compose one data structure. In particular, they are passed by value as one element (Fig. 10).

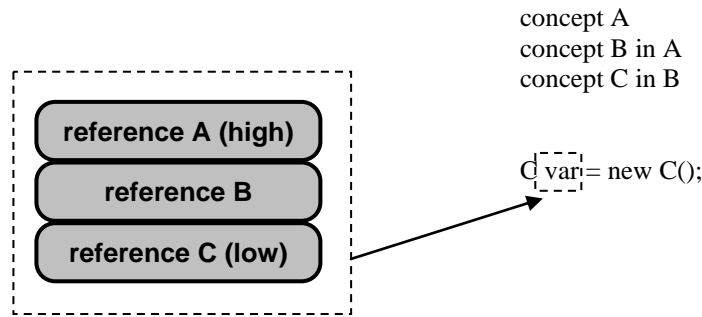

Figure 10. The structure of complex reference.

References have a structure analogous to the structure of objects in OOP. The only difference is that they are passed by value and that the first segment may vary, i.e., it is not necessarily the root of the inclusion hierarchy just like an address does not need to include all the constituents (if they are well known or can be reconstructed). This mechanism allows us to transfer not only an object custom identifier but also references to the space where this object has been created and exists. The object where this object has been created is referred to as a *context*. In our example, the created object of concept `C` is identified by the last (low) segment of the complex reference. The other two segments store references to the parent contexts where this object has been created. These parent spaces are actually normal objects but they are needed in order to access their child (internal) objects.

The structure of objects resulted from the concept inclusion is different from the structure of objects in OOP. The main difference is that each part of an object (object segment) has a separate reference and individual life cycle. In the above example, even if we create one instance of concept `C`, three separate objects will be created in the general case, each having its own reference. One object will have class `A`, the second object will have class `B` and the third object will have class `C`. The references to these objects will be stored as segments in the complex reference but the objects themselves may be stored separately. In particular, one object can be allocated in memory while other objects can be created on disk or remotely on another computer. Another important difference is that base objects can be reused by the extensions, i.e., an already existing object can be chosen as part of the created object. For example, if we create an instance of `C` then already existing objects of concepts `A` and `B` can be chosen and only object `C` is really created. So the new complex reference will contain references to already existing objects in its first two segments and a new reference as its last segment. Thus any new instance can be created in the context of an already existing object and one context may contain many child objects. An important property of this approach is that objects are living in a hierarchy at run time modelled by the concept inclusion relation at compile time.

In CoP each program is based on some ORA facilities provided by default by the compiler and the execution environment. It is the layer which cannot be changed from the program but which is used by the program in order to describe new layers of organization. In the concept hierarchy this default layer is implemented via root concept which is a parent for any other concept. In most cases there will be a library of standard parent concepts for standard ORA mechanisms. For example, there could be a concept for persistent objects, managed objects, remote objects etc. In this case if we want to make our objects garbage collected then their class or concept has to be included into the corresponding parent concept. Notice that in OOP root class provides basic object functions while in CoP root concept provides also reference format and functions.

The concept hierarchy plays an important role because its structure determines how objects in the program will be represented and accessed. In other words, the format of references and intermediate procedures used to access a target object depend on the position of its concept in the concept inclusion hierarchy. The role of each concept consists in establishing a level of indirection for all its objects. It is quite natural because any environment acts as a border for its internal objects. This border has many different functions including object protection, access control and life-cycle management. What is important, the environment may significantly change the behaviour of the objects themselves, i.e., objects in one space may behave differently than objects in another environment. This indirection is nested and repeats the concept structure at run-time.

## 4.2 Method Interception

Earlier we postulated that reference intercepts all accesses to the represented object. If a reference contains several segments then each of them can provide a specific implementation of a method. Then the question is the method of which segment should be executed? In CoP we use the following principle: *parent segments of a complex reference intercept any access to the represented object*. In other words, higher segments have higher priority in processing incoming access requests. If a method is applied to a complex reference which defines this method for all its segments then the first method to execute will be that of the first (high) segment.

Listing 4 provides an example illustrating this principle where concept C is included in B which in turn is included in A and all three concepts define one method for both the reference class and the object class. A reference to a new object of concept C will consist of three segments (line 22). If the method is applied to this reference (line 23) then according to the above formulated principle the compiler will use the definition provided by the reference class of concept A. This method invocation results in the output shown in line 25.

**Listing 4. Method interception.**

```
401  concept A
402    reference {
403      void myMethod() { print("=== A: myMethod() is called"); }
404      ...
405    }
406    class { ... }
407
408  concept B in A
409    reference {
410      void myMethod() { print("=== B: myMethod() is called"); }
411      ...
412    }
413    class { ... }
414
415  concept C in B
416    reference {
417      void myMethod() { print("=== C: myMethod() is called"); }
418      ...
419    }
420    class { ... }
421
422  C myVar = new C();
423  myVar.myMethod();
424
425  $ === A: myMethod() is called
```

This principle is quite natural and simply reflects the fact that any attempt to enter a scope must be intercepted at the border. Higher segments represent external spaces while lower segments represent internal spaces. In order to reach an element we always start from the external space and then proceed by entering narrower scopes (if we are not already inside). At each border the request is intercepted and the programmer can perform the necessary intermediate operations. (Of course, if such a processing is not needed then the interception can be optimized.) Such a sequence of access effectively means that the only possibility to access an object consists in intersecting the intermediate borders that separate it from the outside world.

If a parent reference method intercepts all requests then how methods of the child segments can be reached? For example, in Listing 4 the reference method of concept A (line 3) prints a message and returns without any other actions. So it appears that the methods of concepts B and C can never be reached. However, it is not so. One situation where methods of lower segments can be called is where the complex reference simply does not include higher segments. Such shorter references can be used if the scope of the object is restricted and well known. The access procedure always starts from the first segment really present in the reference and if it is B then the method of this concept will be executed. In the general case the necessity to call child methods depends on the parent reference method itself. Normally after some intermediate processing, say, security checks or some preparations, the parent reference method wants to proceed by passing control to the next segment in the complex reference. In order to identify a child segment we will use the keyword 'sub'. Thus the keyword 'this' refers to the current segment while the keyword 'sub' refers to the next segment in the complex reference.

Listing 5 shows an example of using this mechanism for delegating the current request to the next segment of the complex reference. Here each reference method prints a message and then calls the same method of the next segment (of a child concept). Each method call is then wrapped into a nested sequence of its parent reference methods. Thus each parent reference has a possibility to intervene into the processing of requested send to its children.

**Listing 5. Passing control to child segments in a complex reference.**

```
501  concept A
502    reference {
503      void myMethod() { print("=== A: myMethod() is called"); sub.myMethod(); }
504      ...
505    }
506    class { ... }
507
508  concept B in A
509    reference {
510      void myMethod() { print("=== B: myMethod() is called"); sub.myMethod(); }
511      ...
512    }
513    class { ... }
514
515  concept C in B
516    reference {
517      void myMethod() { print("=== C: myMethod() is called"); sub.myMethod(); }
518      ...
519    }
520    class { ... }
521
522  C myVar = new C();
523  myVar.myMethod();
524
525  $ === A: myMethod() is called
526  $ === B: myMethod() is called
527  $ === C: myMethod() is called
```

In fact, this principle is a dual form of the mechanism of method overriding in OOP, which allows a sub-class to provide a specific implementation of a method defined in its super-class. In other words, we say that reference methods of parent concepts override the corresponding reference methods of the child concepts. (Notice that this principle belongs to reference methods only.) If the reference class of the parent concept does not define the method then it will not be intercepted and then the child method will be called directly. For example, if concept A will not define its reference method then the method of the next segment will be called (in this case B). Alternatively, we can assume that if a reference does not define a method then its definition is provided by default. We say that parent reference methods have precedence over child reference methods. This sequence of access is shown in Fig. 11 where high segment intercepts accesses to the object.

Object methods can be called as usual from any reference method as was described earlier. However, if an object method has been called then it is possible that it is defined not only in this concept but also in parent concepts and hence there is some ambiguity concerning what definition to use. Here we use the conventional OOP principle which says that child object methods have precedence over parent object methods. Informally this means that if we need some service or support from an object then the most specific one will be provided first.

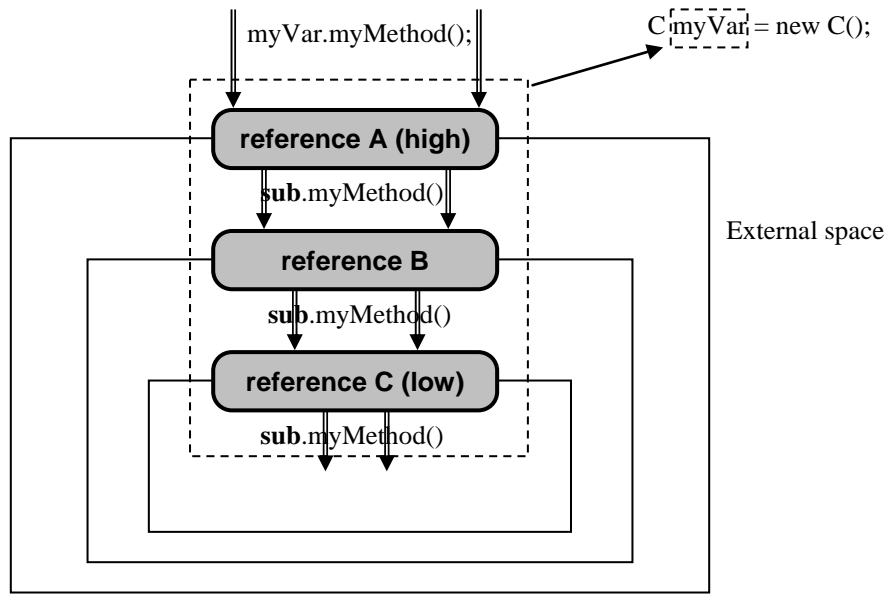

Figure 11. A sequence of execution of reference methods.

An example in Listing 6 illustrates this principle. Here the reference method of the most specific concept C calls its object method (line 23). The object method then delegates this request to its parent object denoted by the keyword 'super' (line 27). The parent object in turn delegates this request to its own parent (line 17). The output of this program is shown in lines 34-39. Here we see that child objects intercept calls to parent objects. The keyword 'super' is used to access the current context, i.e., the object where this object exists. Notice that 'super' denotes an object while 'this' and 'sub' denote references. This means that if a method is applied to 'super' then it is an object method while if it is applied to 'sub' and 'this' then it is a reference method.

**Listing 6. Overriding reference methods.**

```
601  concept A
602    reference {
603      void myMethod() { print("=== A: myMethod() is called"); sub.myMethod(); }
604      ...
605    }
606    class {
607      void myMethod() {print("--- A: myMethod() is called"); super.myMethod();}
608      ...
609    }
610
611  concept B in A
612    reference {
613      void myMethod() { print("=== B: myMethod() is called"); sub.myMethod(); }
614      ...
615    }
616    class {
617      void myMethod() {print("--- B: myMethod() is called"); super.myMethod();}
618      ...
619    }
620
621  concept C in B
622    reference {
623      void myMethod() { print("=== C: myMethod() is called"); myMethod(); }
624      ...
625    }
626    class {
627      void myMethod() {print("--- C: myMethod() is called"); super.myMethod();}
628      ...
629    }
630
631  C myVar = new C();
632  myVar.myMethod();
```

```
633
634  $ === A: myMethod() is called
635  $ === B: myMethod() is called
636  $ === C: myMethod() is called
637  $ --- C: myMethod() is called
638  $ --- B: myMethod() is called
639  $ --- A: myMethod() is called
```

The diagram illustrating this principle is shown in Fig. 12 the lower part of which extends Fig. 11. Here object method calls are shown by single arrows. Each object method call is made using keyword 'super' which allows us to leave the current context and refer to the parent context. Of course, we can call any methods of the current context and not only myMethod as shown in Listing 6. We used one method defined in different concepts in order to demonstrate that the only way to exit the current space consists in crossing the external space border. So each child object protects access to its parent objects just like parent references protect access to child references. This diagram also shows how three keywords are used to refer to different parts of the object and reference.

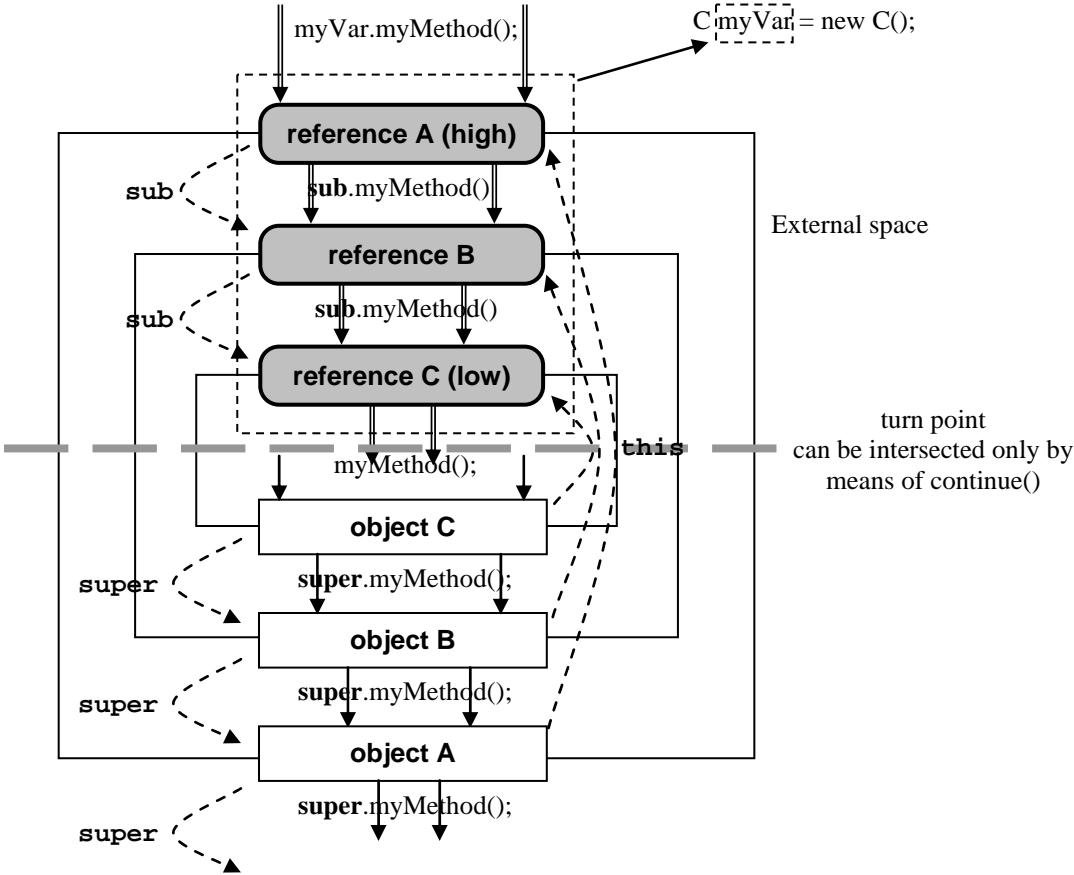

Figure 12. A sequence of execution of reference methods and object methods.

## 4.3 Complex Reference Resolution

When an object of some concept needs to be accessed its reference has to be resolved by means of its continuation method because it is the only way to restore the native reference. This transfer from the realm of references to the realm of objects is called a turn point. If an object is represented by its complex reference then each part of this object has its own reference represented by one segment. Consequently, in order to access some part of this object it is necessary to resolve its segment in the complex reference. If an object is resolved only when it is going to be accessed and only for one access then such a strategy is called *resolution on demand*. For example, if some reference method does not use its object and parent objects at all then no resolution is required. However, if some method accesses this object or parent objects several times then for each such access the object has to be resolved.

Such an approach is not only inefficient but also not very natural. Indeed, assume that we need to get some support from several objects located within some space. Say, we might want to arrange some formalities in a department by signing documents and making agreements. For this purpose we do not enter this space one time for each individual operation, i.e., we do not exit the building and then again enter it in order to do something in another office within this same building. Instead, we enter the space once and then get direct access to all its internal services. The same strategy is used in the concept-oriented programming by resolving complex references in advance *before* any object is accessed. In other words, if the compiler sees that the object represented by the reference is going to be accessed then it resolves the necessary segments of this reference and then all operations can be executed directly using native references.

Since reference segments of the complex reference are resolved in advance the result of this resolution has to be somewhere stored. The data structure that is used for storing native references pointing to parts of the object being currently accessed is referred to as a *context stack*. The sequence of resolution is quite natural. It starts from resolving the very first (high) segment and then proceeds by resolving the next segments till the last (low) segment. This sequence guarantees that all parent contexts are already resolved and ready for direct access when an internal element is reached. Thus methods of child references and objects can directly access their parent contexts using their native references from the context stack.

Let us consider the sequence of complex reference resolution in details (Fig. 13). In general this procedure is analogous to the sequence of reference method invocation shown in Fig. 11. The difference is that the continuation method is used automatically instead of the manual invocation of the business method by the programmer. Another important difference is that the result of each continuation method in the sequence is stored on top of the context stack. (As we already noticed this result is not returned but can be identified by the compiler in the source code.) The sequence starts from resolving the first (high) segment of the complex reference. The native reference restored by the continuation method is stored as the first element on the context stack. Then the procedure starts resolving the next segment by calling its continuation method and again the result is stored on the context stack as the next element pointing to the second object of concept B and so on till the last segment. When all segments are resolved the context stack contains native references to all parts of the object. After that the business method called by the programmer can be started. Notice however, that now all contexts can be accessed directly without reference resolution.

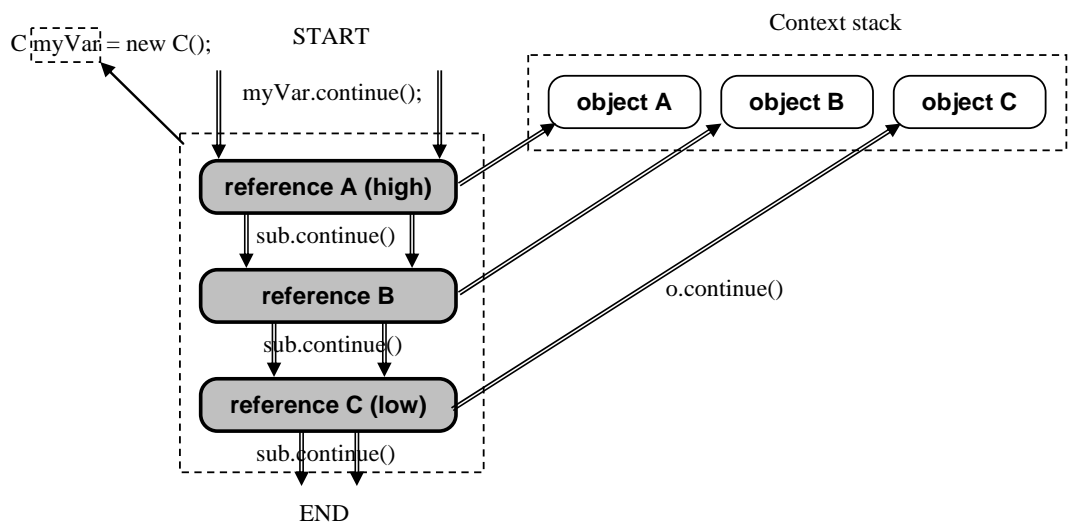

Figure 13. Complex reference resolution and the structure of context stack.

The reference resolution procedure is executed behind the scenes and does not change the main logic of the program. In other words, independent of whether context stack is used or not the program will not change its behaviour. The programmer simply writes the continuation methods and then uses objects as usual by invoking their methods. The compiler is then responsible for possible optimizations of access, particularly, by using context stack.

Let us consider an example shown in Listing 7. Here concept `Account` is included in concept `Bank` assuming that there are several banks each having many accounts. We do not want to deal with the peculiarities of bank account access because they can be stored in a database or even in the bank itself and these operations may require rather complex operations. Therefore we represent banks by their names (line 4) and accounts by their numbers (line 26) in the reference classes. An account reference then consists of two segments which provide indirect access to the object independent of how and where this object really resides. Each of these two segments is being resolved individually by the corresponding continuation method. Bank names are resolved (line 8) using static information loaded at start up (line 1). This procedure is analogous to entering a bank scope and here we could also authenticate the user. Once a bank is entered it is natural that its services can be used without additional resolution and hence the compiler should store the resolved reference on the top of the context stack. Compiler finds the native reference that has to be pushed on the stack by searching for the continuation point (line 9). It is precisely the point where the compiler will start the resolution of the next segment. In this example it is an account number which is resolved by checking that this account number really exists (line 30) then loading its state using the bank internal service (line 31) then starting a transaction again using the bank method (line 32) and finally passing control further to the next method (line 33). At this point the resolved reference is pushed on the context stack because it represents the account object. If this complex reference had more segments (say, a sub-account object) then the next method would be continuation method that would resolve the next object. If it is the last segment then the resolution is finished and the context stack contains all the resolved segments that can be used for direct access. In this example it is important that numerous calls of methods of the parent concept from concept `Account` will be executed directly without resolution. The native reference to the current bank will be taken from the context stack. In particular, if we call some account method (line 50) then the resolution will be carried out only one, i.e., we enter the bank then the account and after that all their services (object methods) can be used from inside directly using keyword 'super' — the continuation methods will not be called again.

**Listing 7. Complex reference resolution.**

```
701  static Map banks = loadBankInfo(); // Load resolution info at start up
702  concept Bank
703    reference {
704      String bankName;
705
706      void continue() {
707        print("> Bank: Start resolving bank " + bankName);
708        Object o = banks.get(this.bankName);
709        o.continue();
710        print("< Bank: End resolving bank " + bankName);
711      }
712      ...
713    }
714    class {
715      ...
716      // Operations within one bank context
717      boolean exists(String key) { ... } // Check if there is such account
718      Object load(String key) { ... } // Load account object
719      void store(String key, Object o) { ... } // Store account object
720      void beginTransaction(String key) { ... }
721      void endTransaction (String key) { ... }
722    }
723
724  concept Account in Bank
725    reference {
726      String accountNumber; // Object identifier
727
728      void continue() {
729        print("  > Account: Start resolving account " + accountNumber);
730        if( !super.exists(this.accountNumber) ) return; // No continuation
731        Object o = super.load(this.accountNumber);
732        super.beginTransaction(this.accountNumber);
733        o.continue();
734        super.endTransaction(this.accountNumber);
735        super.store(this.accountNumber, o);
736        print("  < Account: End resolving account " + accountNumber);
```

```
737        }
738        ...
739      }
740      class {
741        double balance = 0;
742        double getBalance() {
743          print("  --- Account: getBalance is called");
744          return balance;
745        }
746        ...
747      }
748
749  Account account = getAccount("Alexandr Savinov");
750  double balance = account.getBalance();
751
752  $ > Bank: Start resolving bank MyBank
753  $   > Account: Start resolving account 123456789
754  $   --- Account: getBalance is called
755  $   < Account: End resolving account 123456789
756  $ < Bank: End resolving bank MyBank
```

# 5   Operations with References

References in CoP may have different length and each segment belongs to some reference class. This sequence of segments determines position of the object in the virtual address space and the type of processing for each access request. Since references in CoP are of primary importance it is necessary to have adequate means for their manipulation and processing. In particular, it is necessary to have functions for determining a concept for one or another segment. The following three operators instanceof(), contextof() and conceptof() return a concept name given a reference as an argument. This concept name is actually that of one of its segments.

To get a concept name of one segment of a reference we can use operator instanceof() where argument is the reference segment. Let us assume that reference $a = \langle a_1, a_2, \ldots, a_n \rangle$ consists of $n$ segments. Then instanceof($a_i$) returns the concept name of its $i$-th segment. If this operator is applied to the whole reference consisting of many segments then it returns the concept name of the very last segment:

$$\text{instanceof}(a) = \text{instanceof}(a_n), \text{ where } a = \langle a_1, a_2, \ldots, a_n \rangle$$

In a programming language this operator returns the *real* type of any variable according to the reference which is stored in it. Notice that this type can differ from the declaration of this variable and may change in time because the variable may store references of different types (which are sub-concepts of its type).

References do not need to start from the root segment and in this case it is assumed that some higher segments are missing, i.e., the equality instanceof($a_1$) = Root needs not to hold. In other words, a reference may include only local address relative to some context which is not stored in it. In order to get the type of context of a reference the operator contextof() can be used. This operator returns the type of the parent of the very first segment of this reference:

$$\text{contextof}(a) = \text{instanceof}(a_1.\text{super}), \text{ where } a = \langle a_1, a_2, \ldots, a_n \rangle$$

If a reference includes all segments then this operator returns Root as its context. This means that it is a global reference with full context. It is apparent that context is always a super-concept of the real reference type:

$$\text{context}(a) > \text{instanceof}(a)$$

The result of the operators instanceof() and contextof() depends on the real composition of their arguments. In programming languages references are stored in variables which are declared using some type. This type is actually the minimum number of segments this reference must have. In order to return this type operator conceptof() can be used: In other words, if a variable has been declared as having concept `MyConcept` then operator conceptof() will always return this concept name independent of what real reference is stored in this variable. However, it will always be a super-concept of the real reference type:

conceptof(*a*) ≥ instanceof(*a*)

Reference context has to be a super-concept of its declared type. Hence we get the following dependence:

context(*a*) > conceptof(*a*) ≥ instanceof(*a*)

The structure of reference according to these operators is shown in Fig. 14. A reference can be broken into three parts:

- Context is absent in the reference (but must be restored if the object represented by this reference has to be accessed). The type of the last segment of the context is returned by the operator contextof().

- Head of the reference starts from its first segment and ends with the segment having the type as declared in the variable storing it. The type of the last segment of the head is returned by the operator conceptof().

- Tail of the reference depends on the real type of the reference stored in a variable. It consists of all segments that follow the type of this variable as declared. The type of the last segment of the tail is returned by the operator instanceof().

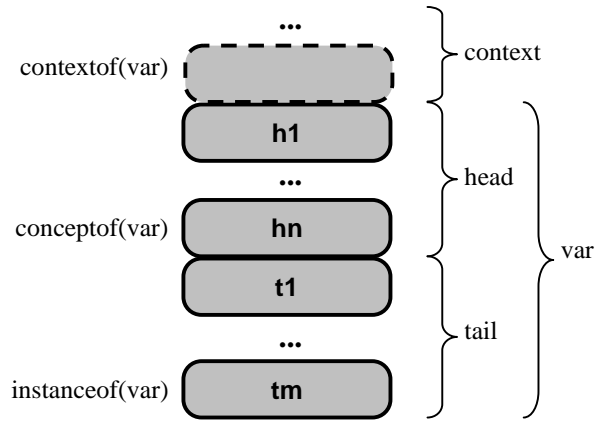

Figure 14. Reference structure and operators.

An available reference may start from one segment while we need it to have another type as the first segment. In other words, we might want to change the context of the reference either by adding new starting segments as a prefix or by removing available starting segments. This operation is called *left casting* of the reference and is written as follows:

LeftConcept : *a*

Here the new desirable type of context is written on the left of the variable. If new segments have to be added then they are taken from the implicit context of this reference. If the reference is shortened then some starting segments are simply removed. The reference returned by this operation has the context specified in the argument:

contextof(LeftConcept : *a*) = LeftConcept

where contextof(*a*) ≤ LeftConcept or contextof(*a*) ≥ LeftConcept

One important use of left casting consists in converting reference to a global reference by adding to it the maximal context:

globalVar = Root:var;

This operation is used when it is necessary to pass a reference that has been used locally outside this scope where it needs to have its context segments to be unambiguous. In other words, it is a method of getting fully qualified reference. Left casting can be also used to extract tail of the reference by cutting off its head part:

tail = conceptof(var):var;

The operation of *right casting* changes the real type of this reference by either removing some last segments or by adding new (empty) last segments:

$a$ : RightConcept

This operation results in a new reference with the real type equal to the specified concept name:

instanceof($a$ : RightConcept) = RightConcept

where instanceof($a$) ≤ RightConcept or instanceof($a$) ≥ RightConcept

Right casting can be used to extract head of the reference by cutting off its tail:

head = var:conceptof(var);

In this way it is also possible to get a reference to any intermediate object represented by this reference (parent context).

Given two references $a = \langle a_1, a_2, \ldots, a_n \rangle$ and $b = \langle b_1, b_2, \ldots, b_m \rangle$ it is possible to find their intersection and union. Intersection of two references includes their common segments:

$c = a \cap b \quad \Leftrightarrow \quad \forall c_k \ \exists i, j :$
conceptof($c_k$) = conceptof($a_i$) and conceptof($c_k$) = conceptof($b_j$)

Union is defined dually as a new reference which contains segments of the both references:

$c = a \cup b \quad \Leftrightarrow \quad \forall c_k \ \exists i, j :$
conceptof($c_k$) = conceptof($a_i$) or conceptof($c_k$) = conceptof($b_j$)

These operations are defined only for the case where the arguments and the result contain a well formed reference, which means that the concept of each segment is a super-concept of the next segment.

Reference concatenation is equivalent to reference union where the real type of the first reference is equal to the context of the second reference. It allows us to get the first reference as the context and then attach to it segments of the second reference as an extension. Concatenation can be defined as right casting the first reference to the type of the second and then copying the second reference to these extended segments:

$c = a : b \quad \Leftrightarrow \quad c = a : \text{instanceof}(b) = b$

Assignment operation consists in copying intersection of two references into the target reference:

$a = b \quad \Leftrightarrow \quad a_i = b_j, \ \forall i, j : \text{conceptof}(a_i) = \text{conceptof}(b_j)$

If their intersection is empty then the target reference does not change. If it is necessary to copy only part of the source reference then it can be cast from left or right:

$a$ = LeftConcept : $b$ : RightConcept

In this case only the segments between LeftConcept (not included) and RightConcept (included) are copied (if they are present in the target reference).

In programming we do not always need fully qualified references with global context which start from the root. In many cases when the context is known or can be restored it is enough to store only the local part of the reference. This corresponds to specifying only the local address in mails which are known in advance to be used in the local context, say, in this country or in this organization. Such short references take less space and are faster during access because the context needs not to be resolved. In order to declare a short (local) reference it is necessary to specify its new context. This can be done by prefixing the variable type. For example, if the allocated variable is of type `MyConcept` but has to contain a reference with the context `MyContext` then we write it as follows:

```
MyContext:MyConcept shortVar;
```

This variable will as usual represent an instance of concept `MyConcept` but its higher segments will not be stored. By default, if the context is not specified then the variable is supposed to have the global context, i.e., it starts from the root:

```
Root:MyConcept globalVar;
```

Given a short reference the question is how to access the represented object if some information is missing. The simplest approach consists in using the current context if it is absent in the reference itself. Notice that the current context is reused rather than resolved again for this reference. This is why access on such references is faster.

The second approach to restoring the context consists in specifying it when a new variable is declared. In this case we provide a concrete variable with context instead of its type:

```
MyContext contextVar = getContext();
contextVar:MyConcept shortVar;
```

Here new variable `shortVar` will be allocated using only segments after context `MyContext`. However, the difference is that the context itself is specified via a variable with concrete value rather than as a concept name. Thus this variable will have a concrete context rather than the current context.

The third approach consists in declaring some context type and then explicitly specifying context only when this variable is used. This approach is useful if there is one context variable and many variables storing local references. By attaching these local references to the context we can produce fully qualified references that can be used to access the objects. This is done by concatenating the context variable and the local references before the reference is used:

```
MyContext contextVar = getContext();
MyContext:MyConcept localVar = getReference();
contextVar:localVar.someMethod();
localVar = getReference();
contextVar:localVar.someMethod();
```

Here we declare local reference `localVar` and get its value from method `getReference`. After that we use this reference to access the represented object by attaching its segments to the context stored in variable `contextVar`. Such an approach is more efficient than manipulating directly global references because not all segments are passed and processed and the context is attached only at the last step when the object needs to be really accessed. This approach can be even used to concatenate several contexts to produce a fully qualified reference:

```
globalcontext:localContext:localVar.someMethod();
```

Here we store global context and local context in different variables which are then concatenated with local object address. Notice that the operation of concatenation does not require precise match between variables – it is smart enough to produce a reference even if the concatenated variables overlap.

Another use of contexts consists in specifying a scope for a block of code. The problem here is that normally the current context is defined by the current object and all computations are executed in its context. In many cases however we need to carry out some actions in the context of another object. The legal way for doing that consists in defining a method of this object class. However, if we do not want to define a special method (for example, if this sequence of actions is used only once) or cannot define a method (if this concept is not available) then the only solution consists in executing these actions in the source context. This might be very inefficient because each action requires resolution of the target reference. In order to overcome this difficulty we can change the context for the block of actions as follows:

```
MyContext myContext = getContext();
myContext : { /* block of code */ }
```

Before this block of code is entered the context `myContext` is resolved just as it is done for executing any method applied to this variable. After that the whole block has a direct access to the context object. Effectively, the block of code is equivalent to a method of the context and the only difference is that it is defined outside the target concept. In particular, all operations in the block are applied to its context by default:

```
MyContext myContext = getContext();
```

```
    myContext : {
      someMethod(); // = myContext.someMethod()
      double b = someField; // = myContext.someField
      :localVar.action(); // = myContext:localVar.action()
    }
```

Notice again that here the target context reference `myContext` is resolved only once when the block is entered and then all operations are executed using direct access.

## 6 Life-Cycle Management

To demonstrate how objects are represented and accessed we assumed that they already exist. However, any object starts its life-cycle from the creation procedure and ends it with the deletion procedure. In OOP object creation is associated with its class constructor – a special method called automatically for initializing the created object. Notice that constructor only initializes the object while its reference is allocated by the system procedure.

In CoP we assume that object creation means first of all initialization of its reference the presence of which reflects the fact of object existence. Once a new reference has been initialized it is assumed that the object can be accessed and hence it is possible to call its initialization procedure. Thus we distinguish two initialization methods: one for initializing references and one for initializing objects. These two procedures are implemented as dual methods of concept, called `create` (Listing 8, lines 5-10 and 16-20). These methods do not return any value but simply initialize the fields of the reference class and the object class. To create a new instance of a concept we declare a variable (line 22) and then apply the creation method (line 23). Here we apply the same sequence as for any other method, i.e., the first method to execute is that of the reference class. It initializes the account number (line 7) and then calls the object creation method (line 8). The object creation method is analogous to the conventional constructor. It initializes the account balance (line 18) and exits.

**Listing 8. Reference and object initialization.**

```
801  concept Account
802    reference {
803      String accountNumber;
804
805      void create() {
806        print("> Account: Initialize reference");
807        accountNumber = getUniqueString();
808        create();
809        print("< Account: Initialize reference");
810      }
811      ...
812    }
813    class {
814      double balance = 0;
815      ...
816      void create() {
817        print("> Account: Initialize object");
818        balance = 0;
819        print("< Account: Initialize object");
820      }
821
822  Account account; // Allocate
823  account.create(); // Initialize
```

The procedure of creation is analogous to the sequence of continuation shown in Fig. 13. The methods of creation and deletion belong to a class of special methods along with the continuation method. However, creation and deletion are normally called explicitly by the programmer while the continuation method is called implicitly whenever the object needs to be accessed.

The creation procedure described above will actually not work and its main purpose is to illustrate the necessity to have two separate methods for reference and object initialization. The main problem is that the reference has been initialized without the object to be actually created. As a consequence it cannot be accessed because the initialized reference cannot be resolved and then the object creation method (constructor) is not able to execute. Thus initializing a reference is not enough to create and access an object. So what does it mean to really create an object? In CoP an object is supposed to be fully created if the whole chain of references up to the native reference has been created rather than only its own reference. In other words, an object is created when its reference continuation method is able to resolve its reference. The resolution procedure implemented by the reference continuation method maps the current reference to a native reference (possibly recursively). Consequently, the creation procedure has to associate the current reference with some base reference. The standard sequence of creation in this case consists in (i) initializing this reference, (ii) creating a new base reference and then (iii) storing this pair in such a way that it can be restored by the continuation method during access.

An example in Listing 9 extends the previous example. Here the creation method initializes this reference (line 15), then creates the native reference (line 16) and after initializing the created object (line 17) stores this pair so that this reference can be resolved in future (line 18). Now we can see that the continuation method of this reference class will be able to resolve this reference by loading the state of the object (line 7). In addition the object creation method (line 17) will be able to initialize the new object because its native reference has just been created and hence the object is accessible.

**Listing 9. Allocating base reference.**

```
901  concept Account
902    reference {
903      String accountNumber;
904
905      void continue() {
906        print("> Account: Start resolving account " + accountNumber);
907        Object o = load(this.accountNumber);
908        o.continue();
909        store(this.accountNumber, o);
910        print("< Account: End resolving account " + accountNumber);
911      }
912
913      void create() {
914        print("> Account: Initialize reference");
915        accountNumber = getUniqueString();
916        Object o.create(); // Initialize base reference
917        create(); // Call this object constructor
918        store(this.accountNumber, o); // Remember for future use in continue()
919        print("< Account: Initialize reference");
920      }
921      ...
922    }
923    class {
924      double balance = 0;
925      ...
926      void create() { // Object constructor
927        print("> Account: Initialize object");
928        balance = 0;
929        print("< Account: Initialize object");
930      }
931
932  Account account; // Allocate
933  account.create(); // Initialize
934
935  $ > Account: Initialize reference
936  $ > Account: Initialize object
937  $ < Account: Initialize object
938  $ < Account: Initialize reference
```

If the created object is represented by a complex reference consisting of more than one segment then the standard sequence of access is executed. Namely, the process starts from the creation method of the very first (high) segment. It initializes this segment and then calls the creation method of the next segment and so on down to the last (low) segment. Such a sequence reflects the fact that the context must already exist before any object within it can be created. In other words, no object can exist without context. For example, if we want to create a new house (number) then we must to know the street where it is going to exist before the house reference can be initialized.

An important point in the creation sequence is that many local references can exist within one and the same parent reference because objects are living within a hierarchy. Accordingly, when a new object needs to be created it does not mean that *all* the segments of its complex reference have to be also created. If some parent context is designed to contain many objects then it can be reused. This means that when a new object is going to be created the reference to this context (and all its parent references) is initialized with the values that represent an already existing object. In other words, instead of creating a new context for a new object we choose some existing context. In other situations the creation method can choose either to reuse an existing context or to create new. For example, if objects are contained in a hierarchical buffer then a new buffer could be created only when all existing buffers are full.

An example in Listing 10 demonstrates this logic of context reuse during object creation. Concept `Container` is developed to serve as a base class for other concepts, i.e., it expects that it will be used to manage a number of internal objects. There exist several containers which have a limited number of objects. The mapping from the space of container integer identifiers (line 4) to the space of native references representing container objects is stored in the index `map` (line 1). In the general case this information would be stored in the parent object. The method of creation is a signal that somebody needs to create a new object within a container (new or existing). The reference creation method tries to find an existing container which is not yet full (lines 9-13) and initialize this reference with the found container identifier in the case of success (line 11). And only if all containers are full (or there are no containers yet) we really create a new container (line 15). Otherwise, if an existing container has been found, we simply resolve its reference (line 16). The container creation and resolution processes are not shown because they are analogous to the procedure described in the previous example. When an existing or new container is resolved the procedure proceeds as usual by applying to this native reference the creation method (line 17). It is precisely the point where all internal objects will be created in a nested manner. This line is analogous to the standard continuation. In particular, this native reference (resolved or new) will be pushed on the context stack and hence this container will be directly accessible by its child objects during creation. Interestingly, the child creation method does not know if the context (container) is really new or has been reused. Finally, if the container has been really created we store this information in the map for future accesses (line 18). (So before creation is finished nobody can access this new container.)

**Listing 10. Context initialization.**

```
1001    static Map map = new Map();
1002    concept Container
1003      reference {
1004        int containerId;
1005
1006        void create() {
1007          print("> Container: Find or create container");
1008          boolean found = false;
1009          forall(Container c in map) {
1010            if( ! (c.size() > 100) ) {
1011              this.containerId  = c.containerId; found = true; break;
1012            }
1013          }
1014          Object o;
1015          if(found == false) o = createContainer();
1016          else o = resolveContainer(this.containerId);
1017          o.create();
1018          if(found == false) map.add(this.containerId, o);
1019          print("< Container: Find or create container");
1020        }
1021        ...
1022      }
1023      class {
1024      }
```

The actual creation procedure used for one or another reference class can be quite complex and depends on the requirements of the application being developed. For example, base references need not to be really allocated but can be reused. In this case the concept maintains a pool of references from which unused references can be allocated during creation process. For example, a concept might maintain a pool of memory handles or it could allocate one buffer and then manage blocks of memory in it itself. It is also possible to use lazy creation where base references are allocated only when this object is really needed, i.e., when it is actually accessed. For the mechanism of instance creation it is important only that each concept has two methods with special role while their implementation is provided by the programmer.

The logic described above for the process of creation is applied to object deletion. Each concept can define deletion methods for its reference class and object class. The reference deletion method is intended to clean up this reference and destroy all the references it resolves into up to the native reference. It is assumed that after that this reference cannot be used for object access and hence the represented object is considered non-existing. Normally deletion method resolves all the segments and then deletes then consecutively starting from the last (low) and ending with the first (high).

The object deletion method is analogous to the conventional destructor and its main purpose consists in cleaning up the object state before it is really deleted. Actually the object deletion method is the latest chance to do anything with this object. In real applications implementation of the deletion method can be quite complex depending on the mechanism used for object management. In particular, it is possible to avoid explicit object deletion at all if some parent concept implements the logic of automatic deletion of unused objects. Deletion does not necessarily results in the real physical removal of all allocated resources such as native reference to a memory location. Instead, these resources might be returned to a pool of unused objects from where they are used when creating new objects. All these implementation details are defined by the programmer while for the compiler it is only important to follow the sequence of deletion which is analogous to all special methods like continuation and creation.

As usual the creation and deletion methods can be defined as taking some initialization parameters. Some segments of a new object reference could be initialized by the programmer. For example, if we have already a reference to a container and want to create a new object in it then it could be done as follows:

```
Container container = getContainer();
Account = account = container;
account.create();
```

Here we explicitly assign the container segment and then the creation method should check its value. If it is null then the container will be either created or chosen from existing. If it is not null then this value will used as a reference to the desired container for a new account.

# 7 Inheritance and Polymorphism

In OOP the mechanism of inheritance is indented for extending behaviour of an already existing objct class. In particular, it is possible to add new fields and new methods by producing a more specific class reusing the functionality of the base class. In CoP this mechanism still exists but is made more general and the classical interpretation is only a particular case of the concept-oriented inheritance. The main distinction is that each new object class extension gets much more independence from its base class. Essentially, each part (segment) of a complex object described by one object class has its own representation (reference) and hence its own life-cycle. In contrast, in OOP all object segments live together as inseparable parts of one complex object which starts from the base segment and ends with the last extension. It is not possible to create/delete the individual segments or to manage them separately. Thus inheritance in CoP means *inclusion* in a parent object or assigning a context which can be shared among many other objects. The base segment (context) is then a common part of many extensions rather than belongs to this and only this object. For example, if we have base type `Panel` and its extension `Control` then in OOP creating one control means also creating one panel for it. This new panel consists of two object segments represented by one native reference (Fig. 15). In CoP panel objects can be reused, i.e., many controls can be created in the context of one panel. Thus the extensions live separately and have their own individual references which are local identifiers in the context of one base panel object.

Another important distinction from OOP is that references in CoP intercept any accesses to this and child objects. Such a sequence of access significantly changes the mechanism of polymorphism. In OOP polymorphic behaviour has built-in character where it is guaranteed that the object has behaviour of its real class determined at run-time (rather than declared at compile-time). How this semantics is implemented is not known to the programmer. In CoP such built-in standard semantics is absent and polymorphic behaviour is replaced by the special sequence of access. The main idea is that each base reference intercepts all calls to child references and objects. Hence it can decide at run-time how to proceed and what to do. In particular, if there are no child elements then the base reference implements the logic of the base concept normally using functions of the represented base object. In the case there is a child object, the base reference can implement some intermediate logic and then pass the request to the child reference. This child again can check if it is the last element and then decide how to further process this request: either to pass it to the next element or to finish execution. For example, let us assume that there is base class `Figure` extended by class `Rectangle`. If we declare a variable of the base class which is then assigned a reference to a rectangle object then method `draw` applied to this variable will draw a rectangle. In OOP this is done automatically by definition of polymorphic behaviour. In CoP, the base reference will intercept the invocation of the method `draw`. Then it can check if there is an extension and decide how to proceed. In the case of the rectangle object the base reference can simply pass this request further and the rectangle reference will be responsible for drawing this object of class `Rectangle`. In other cases the base reference may do something specific to the current level and only after that pass the request to the child element. For example, the reference and object of concept `Figure` might draw a background.

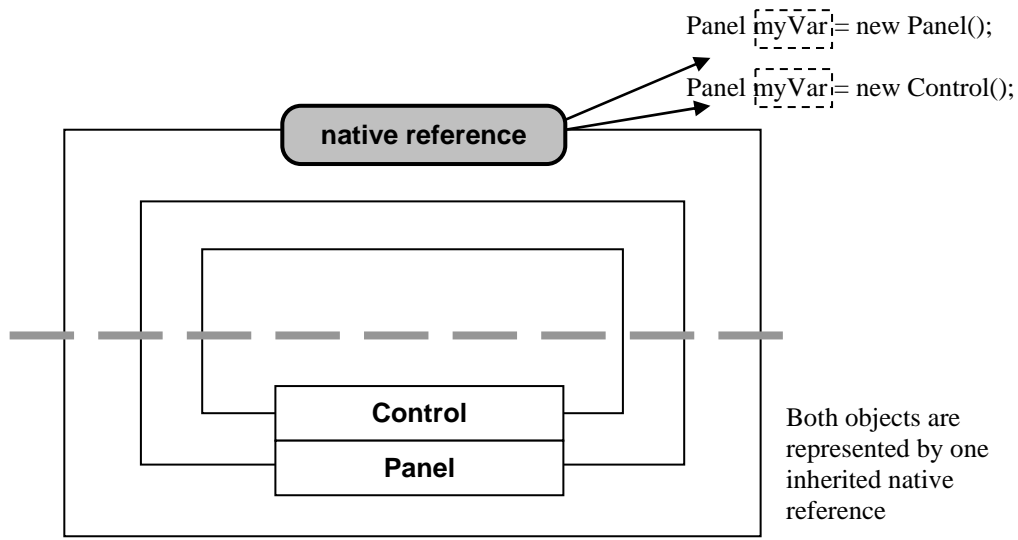

Figure 15. All segments of a complex object are represented by one parent native reference in OOP.

Let us demonstrate how polymorphism works in CoP using an example shown in Listing 11. Base concept describing a bank account (line 1) implements method `getBalance` in both the reference class and the object class. Objects of the concept `Account` are represented by text strings (line 3) which are resolved at run-time as described in the previous sections. Now let us assume that this concept describes only the basic behaviour of a generic account and is supposed to be extended by concepts describing concrete account types. In particular, we might define concept `SavingsAccount` which has more specific functionality by including it in the base concept `Account` (line 20). Concept `SavingsAccount` has its own reference which however does not have any fields. Effectively, this means that for each base object there will be one extension object both represented by one base reference (account number). Therefore the object structure is the same as in OOP except that the base object has its own custom reference (in OOP only native references are possible). Yet the reference class of concept `SavingsAccount` defines method `getBalance` (line 22) which plays an important role in implementing polymorphic behaviour. The object class of concept `SavingsAccount` defines the account balance field (line 32) already existing in the base class. Thus an instance of this concept will have actually two balance fields: one in the base object and one in the extension. In this example it may be not so important because there is one-to-one correspondence between base objects and extensions. Yet if one base account could have many extensions then this would make more sense. The object classes of the both concepts define method `getBalance` as returning the current balance of either the base account or the savings account.

If a variable is declared as having the base type then just as in OOP it may contain a reference to an object of any its extension. However, in OOP this reference is always of native type and the real object class has to be somehow restored at run-time, normally by using an additional field in the object (for example, pointing to a table of virtual functions). In contrast, in CoP the reference may have a different number of segments. Each additional segment in the reference means that it points to an extended object. For example, if we declare a variable as having type `Account` and store a reference to an object of type `Account` (line 37) then the last segment of this reference will be of concept `Account`. In this case, if we apply method `getBalance` to this variable (line 38) then its implementation in the reference class of concept `Account` will be executed (line 5) because it is the very first segment of the reference. This method checks the real (run-time) type of this reference and then chooses the continuation path. Actually it is important only if this reference is the last segment or not, i.e., if there exists an extension or it is the last object. If it is the last segment we simply call this object method (line 7), which returns the account balance. Otherwise, if there is an extension we simply pass this method to the next segment (line 8). Notice that here we do not know what is the type of extension and what kind of processing it will do. The base reference method simply performs some actions and then makes it possible for the children to contribute to the processing of the request. In this example, the savings account reference will also check whether it has the last segment and then either return the balance of this object (line 24) or continue processing further by passing the request to the next child (line 25).

**Listing 11. An example of polymorphic behaviour.**

```
1101  concept Account
1102    reference {
1103      String accountNumber;
1104
1105      double getBalance() {
1106        print("> Account: Start interception " + accountNumber);
1107        if( sub == null) getBalance(); // This object method
1108        else sub.getBalance(); // Forward to the subobject
1109        print("< Account: End interception " + accountNumber);
1110      }
1111
1112      ...
1113    }
1114    class {
1115      double balance = 0;
1116      ...
1117      double getBalance() { return balance; }
1118    }
1119
1120  concept SavingsAccount
1121    reference {
1122      double getBalance() {
1123        print("> SavingsAccount: Start interception " + super.accountNumber);
1124        if( sub == null) getBalance(); // This object method
1125        else sub.getBalance(); // Forward to the subobject
1126        print("< SavingsAccount: End interception " + super.accountNumber);
1127      }
1128
1129      ...
1130    }
1131    class {
1132      double balance = 0;
1133      ...
1134      double getBalance() { return balance; }
1135    }
1136
1137  Account account = getAccount("Alexandr Savinov");
1138  balance = account.getBalance(); // Account::getBalance
1139
1140  Account savingsAccount = getSavingsAccount("Alexandr Savinov");
1141  balance = savingsAccount.getBalance();// SavingsAccount::getBalance
```

From this example we see that a variable may have a base type and the same method applied to the stored reference results in different sequences of actions depending on the real object type. However, polymorphism in CoP works differently in comparison with OOP. In OOP, polymorphism is much simpler and is reduced to calling the method defined in the real object class. In CoP, polymorphism is a processing chain consisting of several methods. In the general case, each intermediate reference and object may contribute to the processing of any request. The method executed by default in OOP is only the last step of the chain of request processing in CoP. Such a view of polymorphism is a particular case of the general concept-oriented principle that method execution is a *sequence* of steps which correspond to the points where the access request intersects intermediate borders on its way to the target object. Interestingly, it is not guaranteed that the last method will be reached at all. For example, the base object may raise an exception for any reason such as security violation or insufficient resources. In OOP we always know that the method of the object that has been called will be executed. An advantage of the concept-oriented polymorphism is its flexibility because the programmer is able to control the whole sequence of access. However, in simple situations it is less efficient because in OOP the indirection used for calling virtual methods is optimized by the compiler.

Objects in CoP are living within a hierarchy which is modelled by the concept inclusion relation. In this case each object segment of a complex object belongs to one context (or base) segment and may have many own extension segments. The object segments are distinguished within one context by their own local reference segments. For example, let us suppose that main account may have several sub-accounts with different purpose and parameters including different savings accounts. In order to model such a case we can add identifying fields in the reference class of concept `SavingsAccount`. The savings account object will be then created separately from the main account. However, their reference will include the main account number as the first segment. Method `getBalance` of the base concept could return the total balance of all the sub-accounts while this method for each sub-concept such as `SavingsAccount` could return balance for this individual account only. In contrast to OOP, here an object may have many extensions at run-time. A method invocation is viewed as a name for the path that has to be followed on the way to the target object represented by its complex reference.

# 8   Dual Methods

Each concept provides two types of methods with the same signature: reference methods and object methods. The role and properties of these methods are significantly different within a hierarchy. Reference methods are used to enter a scope, i.e., they are executed when an access request needs to come into the space of the object represented by the reference. Therefore they can be thought of as incoming methods. An important property of reference methods is that base methods have precedence over the same methods of the extensions. In other words, if a method has been applied to a reference then the base method will intercept this call. The sequence of access for reference methods is shown in Fig. 16, left. Reference methods of extensions can get control only if their base methods will call them explicitly using keyword 'sub'. Thus the conventional principle of overriding object methods has the opposite form for reference methods: *a base reference method overrides its extension reference methods*. This means that it is always possible to override any reference method by defining the same method in the reference class of its base concept. It is quite natural principle because it allows any border to control incoming processes. For example, a live cell has a border which checks and controls anything that tries to come in. The existence of the own border is actually a generic property of any system including physical, live and social ones. And it is quite natural that if a system is included into another system than the only way to access it from outside consists in intersecting the parent system border which controls all the incoming processes. The mechanism of reference methods and the inverse principle of overriding allow us to support this approach in programming languages.

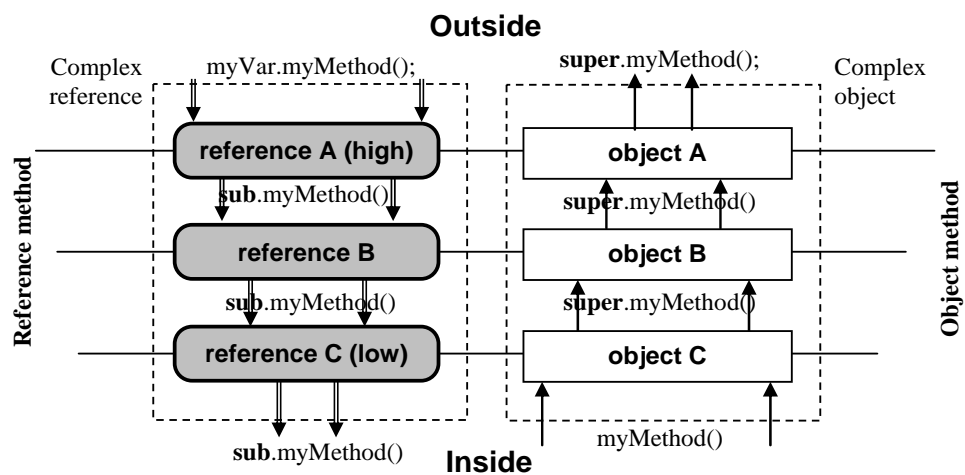

Figure 16. Overriding reference and object methods.

Let us assume that there is concept `Account` with possible sub-concepts describing special account types. Since there can be many different account types at compile time and account objects at run-time we would like to control when and how some method of all those different objects is called. Say, we need to check more thoroughly the conditions under which method `getBalance` is called. Notice again that this method can be called for objects of different concepts but we need one point in the program code through which all they pass. For example, in this point some security checks will be performed and general diagnostic messages written. Since base reference methods override extension methods we can define a new base concept `Diagnostics` with this reference method implemented and then include all other concepts into it (Listing 12). Method `getBalance` in concept `Diagnostics` will intercept all incoming calls to any instance of a concept included into it. In particular, if concept `Account` is included into `Diagnostics` (line 20) then its method `getBalance` will always be intercepted by the parent concept, which prints concept name of the object being accessed (lines 5). So in this example the base reference method overwrote all future extension reference methods. However, instead of complete substitution of method definition in OOP, it actually intercepted method call and after some intermediate actions forwarded the request further to the extension.

**Listing 12. Intercepting incoming and outgoing calls via transparent concept.**

```
1201    concept Disgnostics
1202      reference {
1203        // No fields
1204        double getBalance() {
1205          print("> Disgnostics: getAccount for: " instanceof(sub));
1206          sub.getBalance(); // Forward to the sub-object
1207        }
1208
1209        ...
1210      }
1211      class {
1212        // No fields
1213        double getBalance() {
1214          print("> Disgnostics: getAccount to: " instanceof(super));
1215          super.getBalance(); // Forward to the super-object
1216        }
1217        ...
1218      }
1219
1220    concept Account in Diagnostics ...
```

Dually to reference methods, object methods are intended to be called from inside, i.e., from extensions. It is not possible to access an object method from outside and the only way to use it consists in entering the object scope via some reference method and then calling it from inside. So object methods can be thought of as internal services intended for trusted users who already crossed the border. Object methods build their functionality on the basis of their parent object methods as shown in Fig. 16, right. This means that when an object method is called it can ask for support from its base object which in turn may call its own base object methods and so on. This sequence of access is actually opposite to that followed by reference methods. And the rule of overriding has also the opposite form. Namely, an object extension method has precedence over its base object methods. In other words, an object method intercepts all calls to the same base methods. This effectively protects base object functions from being directly called by extensions and this mechanism allows the programmer to establish control over the use of internal services. For example, concept `Diagnostics` implements method `getBalance` in its object class (line 13) which means that all calls of this method from any extension will be intercepted at this level.

It can be said that reference methods protect internal elements while object methods protect external elements. Another analogy is that reference methods are thought of as callback methods which are supposed to get external messages or event notifications. Dually, object methods could be thought of as callforward methods providing base services to internal elements located in their context. This duality is very important for concept-oriented programming. Frequently we can define concepts without fields just to intercept method calls: either incoming via reference methods or outgoing via object methods. Two types of method overriding (object and reference) play an important role in facilitating polymorphism described in the previous section.

# 9 Related Work

The approach described in this paper relates generally to the methods of indirect object representation and access (ORA). More specifically it relates to the methods of reference modelling in programming languages. The traditional methods and technologies that can be used to solve these problems are listed below in ascending order of the system organization complexity:

- hardware level
- operating system level
- middleware and custom run-time libraries
- frameworks and language infrastructure
- programming patterns
- language support for access indirection
- language support for reference modelling

Access indirection exists at all levels of the system organization including physical interactions and deeper just because instant access is an abstraction and any access requires some intermediate environment and some time to propagate. However, here we start the discussion from the hardware level of the system organization ignoring all the lower level issues. It is frequently assumed that memory addresses provide direct access to its contents. However, it appears so only from the point of view of the programmer while for the processor each such access means quite a lot of work. First of all it is important to understand that (in contemporary architectures) each address is actually a location in a *virtual* address space, i.e., it is an abstract space which is not directly connected with the real memory. This virtual address space is supported by the processor which is responsible for its maintenance. In particular, the processor needs to map each virtual location in a real cell in the available physical memory. Notice that the amount of physical memory may be less than the size of the virtual address space and it might be necessary to have larger storage for swapping some address intervals. The processor has also to check permissions each time a memory location is accessed and this operation also takes some processor internal micro-cycles. The virtual address may have some structure which needs to be mapped to the physical address structure. For example, earlier 16-bit processors manipulated addresses consisting of one 16-bit segment and one 16-bit offset stored in two registers. These two elements of the address produced 20-bit full address by means of 4-bit shift executed by the processor during access. Thus it is important to understand that the conception of indirect access, custom references and address space virtualization actually exists already at the hardware level. The approach described in this paper can be viewed as providing similar means at the level of the programming language.

Processor accepts memory addresses and then resolves them into physical locations in memory. However, using the processor virtual address space is frequently not very convenient for application programmers. Therefore a new level of indirection is normally created by the operating system. Global heap where pieces of memory can be allocated by applications is an example of such a service. Each element in the heap has its own unique identifier which is however not directly connected with the corresponding address in memory. In order to access a piece of memory represented by such an identifier it has to be locked and resolved by the application. Operating system keeps a mapping from the space of memory handles into the memory addresses. In addition, operating system might also provide other types of containers with their own address spaces. For example, for small objects an application might use local heap. It is important that just as for hardware level, the programmer cannot customize these facilities or develop new levels of indirection.

The idea of middleware-based approaches consists in creating special software and hardware environments where a conventional program will run. In particular, such an environment offers a number of functions that are intended to support indirect representation and access functionality. This special environment can exist and be accessible to running programs in very different forms, for example, as part of an operating system, an object container, a service, a dynamically or statically linked library etc. However, the main property of this approach is that the programming language remains the same while the support is provided by developers of the middleware. One wide-spread class of middleware is techniques for remote procedure calls. Examples of such middleware platforms are CORBA and RMI/EJB [Mon06]. These environments provide facilities for creating remote references and then making transparent method calls. As a consequence access to object is even more indirected. Such middleware platforms may fit well to the purposes of one system but may be inappropriate for another system. In other words, it is yet another standard level of indirection which however exists separately from the program and hence cannot be easily adapted to the purposes of each concrete program.

Methods of access indirection can be made part of a programming language infrastructure. An advantage is that the language run-time environment is closer to the programmer and hence it can be better controlled in comparison with the middleware-based approaches. One technology that can be used at this level consists in changing the behavior of a language from this very language. This allows the programmer to adapt this language features to the needs of each concrete program. These possibilities are provided by reflective environments and metaobject protocol [Kic91, Kic93]. Normally programming languages are defined in such a way that their behavior cannot be changed. In particular, we cannot change how objects are represented and accessed because it is hard-coded into the language and its environment. The reflective approach allows the programmer to change this environment and to change the way how the language constructs are interpreted.

Another wide-spread approach to access indirection that belongs to this category consists in providing the mechanism of access interception at the level of the language run-time environment. In Java Virtual Machine this mechanism is called *dynamic proxy* [Blo00]. It exists also in C# where its functionality is implemented in the `RealProxy` class.

Above we described non-language approaches which can be quite useful if indirect access is an auxiliary technique. However, if indirection is regarded as one of the primary mechanisms then it should be supported at the level of the programming language itself. This allows the programmer to express an *arbitrary* logic of indirection without restrictions imposed by the hardware, operating system, middleware or a library. The simplest approach to using a new mechanism in a programming language consists in following some discipline or pattern. One such wide-spread pattern for implementing indirect access in an object-oriented programming language is called *proxy*. Proxy is a special class that emulates the interface of the corresponding target class but inserts some intermediate functionality. (Proxy as a pattern should be distinguished from dynamic proxies as a built-in feature of run-time environment.) These intermediate functions of the proxy class are called before the target methods and hence they effectively intercept all target object method invocations. For example, if we need to indirect access to class `Account` we could define its proxy class `AccountProxy` implementing the same methods. The trick here consists in using proxy class instead of the target class, for example, we need to explicitly use class `AccountProxy` instead of class `Account`. Thus it is not a real interception but rather a normal sequence of method calls using static and explicit substitution. In other words, in the source context a reference to the proxy instance is created and hence the methods of the proxy are called when it is used. Then it is the task of the proxy to decide what to do if some its method has been called. Normally, after some processing the corresponding target method is called. One disadvantage of this approach is that it requires significant manual support and is not very general. It is more a special technique or specific programming pattern rather than a programming paradigm. Here are other disadvantages of this approach:

- If the target class changes then its proxies need to be updated because this pattern is a manual implementation of a discipline while transparency needs to be supported by a mechanism of the programming language.

- One limitation is that a proxy is developed for one target class because it is aware of and explicitly simulates its behaviour. A true interceptor has to be able to intercept method calls to *many* different classes.

- It is difficult to impose behaviour in a nested manner (creating a proxy for a proxy) because it requires even more manual support and such a program is even more sensitive to changes which have to be propagated over the source code.

- This approach allows the programmer to implement indirect access but it does not provide means for modelling references which are passed and stored by value instead of native references. A reference to a proxy is still a normal native reference.

Another interesting programming pattern which is intended to support indirect representation and access requires support from the programming language. In C++ it is based on using templates which are used to implement so called *smart pointers* [Str91]. Smart pointer is an instance of a class that is developed to model references which are passed by value and provide access to an object. For example, we might define class `MyReference` and then instantiate it whenever we need to represent an instance of some target object. Here we already assume that such a class will serve many different target classes by encapsulating general behaviour common to all of them. However, although the reference class has a generic form this solution assumes that we still need to know the target class for the smart pointer to work properly. For this purpose the smart pointer class is parameterized by the name of the target object class using the mechanism of templates. This parameter is then used in implementing methods of the smart pointer class. When a new smart pointer has to be created we must provide the name of the target object class as the value of the parameter, for example, as follows:

```
MyReference<Account> account = new Account();
MyReference<Person> person = new Person();
```

Thus in this approach we must specify both the target class and the reference class for each new instantiation. In particular, an object could be represented by many different reference types. The reference class (smart pointer class) and the object class are connected at the time of instantiation rather than at the time of the definition. The real interception is performed by overloading the universal access operator (dot or arrow in C++). This means that each time a method is applied to a smart pointer this overloaded operator is called and then executes the intermediate actions. Normally it resolves the reference by finding the location of the target object and then calls the target method.

The same idea of having a reference class parameterized by a target object class is implemented in the Transframe programming language [Sha]. The difference is that in Transframe reference classes are supported at the level of the language, i.e., it is a built-in feature while in C++ it is a pattern based on the mechanism of templates. In this language we can mark a class as intended to represent other objects as follows:

```
class MyReference is referential {
  private:
    obj: ObjType;
  public:
    enter(path:char[]) { ... }
}
```

When a new instance of an object has to be created we need to specify both the reference class and the target object class:

```
account: MyReference of Account;
person: MyReference of Person;
```

Thus any declaration of an indirectly represented object requires two classes just as in the case of smart pointers. However, the parameterization is supported by the programming language rather than uses the general mechanism of templates. Below we list main properties of these two approaches and their difference from CoP.

- Smart pointers and reference classes in Transframe use independently declared reference classes and object classes while in CoP these two classes are defined as parts of one more general construct (concept).

- Smart pointers and Transframe assume that a new instance of an object requires two parameters: a reference class name and the object class name. Thus association between these two classes is established at the moment of creation. In CoP the association between the reference class and the object class is established within the concept at the moment of its declaration while creation is performed precisely as in OOP using only the target object class.

- Smart pointers and references in Transframe can be viewed as explicit proxies because they are created as objects of certain class which are then used instead of the target objects. In CoP we have an illusion of working directly with the target object while the indirection is completely hidden.

- Smart pointers and Transframe allow the programmer to implement simple substitutes passed by value but it is difficult to implement hierarchical (complex) references consisting of several segments. And it is even more difficult to implement objects consisting of several separate segments.

- Smart pointers intercept method calls by overloading access operator. In CoP the interception of individual methods is performed by reference methods of the concept which have precedence over object methods and have an opposite overriding direction. General interception can be also implemented via continuation method.

There exist also other language-based approaches to automating indirect access. For example, one general purpose method, called attribute-oriented programming, consists in using annotations or tags to mark up parts of the source code where additional intervention is needed. Another interesting approach, called language-oriented programming (LOP), consists in developing custom languages for each task or problem domain. In LOP the programmer could develop or use libraries of languages just as libraries of procedures or classes are used in procedural programming or object-oriented programming. In particular, such a custom language could support functions of indirect representation and access. Other general approaches that could be used to implement different elements of the mechanism of indirect access include mixins [Bra90, Sma98], subject-oriented programming and multidimensional separation of concerns. These methods allow the programmer to describe how behavioural granules have to be distributed throughout the system. However, they are not targeted at the problem of indirection of representation and access.

One of the most interesting recent approaches to programming is aspect-oriented programming (AOP) [Kic97]. Aspects describe intermediate functionality (and data) injected into the points in the program which are specified by means of regular expressions. Thus aspect can be viewed as a special programming construct that modularize intermediate functionality. An important property of this approach is that aspects know explicitly the points where the intermediate functions will be injected while the target classes do not know what other code will modify their behaviour (Fig. 17). Such a structure of dependencies between the module with the code to be injected and the modules where it has to be injected can be viewed as declaring all the target classes within the aspect (the target classes being unaware of this aspect). In this sense CoP is characterized by the opposite direction of the dependence. Namely, the module with the code to be injected (parent concept) is unaware of the points where it will be used (child concepts). What is similar between AOP and CoP is that method invocation is indirected and can trigger quite complex intermediate actions. However, the principles behind this indirection are quite different.

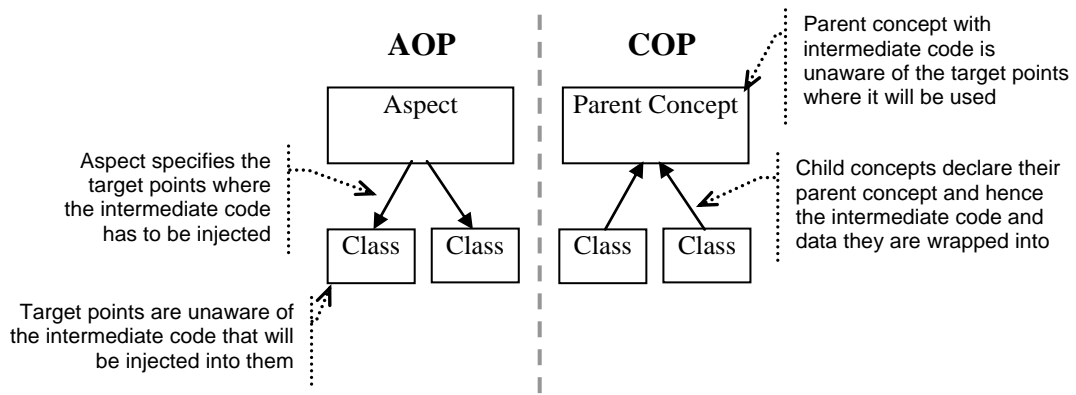

Figure 17. Aspect-oriented programming vs. concept-oriented programming.

The mechanism of dual methods is similar to the approach described in [Gol04] where inner and super methods are discussed. The direction of interception of inner methods is the same as for reference methods. The difference is that inner and super methods belong to an object class, i.e., they are both defined in one class. In CoP two types of methods are defined in two classes within one concept.

There can be two approaches to programming based on using concepts defined as a pair of one reference class and one object class. These approaches depend on the role played by these classes and their responsibility. In papers [Sav05a] one approach, called CoP-I, was described where references of a concept represent objects of its *child* concepts (not this concept). In this paper we have described an approach, called CoP-II, where references represent objects of *this* same concept. Both approaches have some advantages and disadvantages. On earlier stages of development of the concept-oriented programming we thought that the first approach is more perspective. Later on, when trying to remove some subtle problems we gradually switched to the second approach described in this paper.

It should be noticed that there exist other approaches to programming using the term 'concept' for naming their mechanisms or constructs but which do not intersect with CoP. Among such approaches are concept-oriented programming by B. McConnell [Mcc99], concept programming and XL programming language, concept-oriented logic programming [Voi92].

# 10 Concept-Oriented Paradigm

In the paper we proposed a method of generalizing classes by attaching to them a reference class. This construct is called concept and consists of two classes: a reference class and an object class. Both the constituents of concepts have their own members defining their individual behaviour. Concepts are organized into a hierarchy using inclusion relation which generalizes class inheritance. References and complex references are passed-by-value and their main role consists in representing objects. The approach to programming based on using concepts is called the concept-oriented programming.

In describing the concept-oriented programming we have focused manly on technical issues by specifying the roles of different elements. However, this approach is not simply an additional mechanism that can be implemented in existing programming languages. Rather, it has significant influence on the way a complex software system is viewed and a computer program is being developed what is normally called a paradigm shift. In fact, this change of paradigm preceded the development of the described language means, i.e., first, we formulated general principles and then developed several versions of language means to support them. It should be also noticed that this research has been performed in close connection with the corresponding data modelling techniques because they share the main principles. Below in this section we try to outline the general principles of this programming paradigm.

Object is a thing in itself or reality which is not observable in its original form because the realm of objects is radically unknowable. The only way to get some information about an object consists in using its references and hence we always have some intermediate element in computations. In contrast to the reality it represents, a reference is a phenomenon observable directly as is in its original form. Such a separation corresponds to Kant's view expressed originally in his seminal work "Critique of Pure Reason". The separation between the realm of references and the realm of objects plays a fundamental role in the whole concept-oriented paradigm.

One of the main contributions of the concept-oriented paradigm is that it completely legalizes references as a crucial element of any system (while CoP proposes to use concepts for their modelling in a program). References are made first class citizens with the same rights as objects. The shift of paradigm is that having only objects is not enough to efficiently describe behaviour of a system. Such an object-oriented picture is not complete what is especially problematic in complex systems. It can be completed by adding the notion of reference as an object representative. However, the main contribution of the concept-oriented approach here is that references are not simply recognized as an important element but made dual to objects, i.e., they may possess behaviour just as objects. In some sense references can be viewed even more important elements than objects because a system can exist without objects but it cannot exist without references. Moreover, a system consisting of only references may possess rather complex functionality. Another important property is that development of a system starts from developing the structure and functions of its references, i.e., it is more important how objects are represented and accessed while object functions can be defined later.

The duality of references and objects can be considered a continuation of a very general and deep principle of Separation of Concerns formulated by Dijkstra [Dij76]. The main idea of this principle is that any problem or system functionality can be viewed from different points of views or concerns. One specific feature of the concept-oriented paradigm is that we distinguish *two* orthogonal concerns any program consists of: behaviour of references and behaviour of objects. To develop a program functions of references are as important as functions of objects. In this sense it is very important to have adequate and convenient means for modelling inseparable unity of these two concerns and concepts in CoP satisfy most of the necessary criteria.

An amazing feature of concepts is that they smoothly generalize classes and hence the concept-orientation can be viewed as a generalization of the object-orientation which has been a dominant theory for many decades. This generalization is not restricted by the simple fact that concept can be obtained from class by attaching a reference class to it. It covers also such issues as inheritance and polymorphism which take new more general forms within CoP.

Informally, the relationship between concepts and classes is analogous to that between complex numbers and real numbers in mathematics. Like complex numbers consisting of imaginable and real parts, concepts involve two constituents but are manipulated as one construct which may exhibit properties of both references and objects. In the same way as complex numbers are much more expressive and natural for many difficult mathematical tasks, concepts are much more expressive and natural in computer science including programming, data modelling and design.

Traditional approach to programming consists in describing a sequence of actions which can be defined as procedures or class methods. A procedure or method call in this case results in an immediate execution of its operations. Although in reality it is not so and there is always something out there, the existing approaches simply ignore the events happening under the hood. CoP significantly changes this view and here again we see the paradigm shift. In the concept-oriented approach interactions cannot propagate instantly and hence any access requires some intermediate environment for its transfer and additional operations before the target object is really reached. Thus any interaction is actually a *sequence* of operations executed on the way from the current location to the destination specified in the target object reference (Fig. 18). As consequence, the programmer cannot say precisely what happens after a method is applied to an object. A method call is an order or a final goal but its execution depends of the intermediate elements of the system and the position of the target in the address space. For example, crediting a bank account involves one arithmetic operation applied the account object. In OOP calling this method would result in precisely what is written in its definition, i.e., executing the arithmetic operation. In CoP this arithmetic operation will be only the last action in a sequence of possibly hundreds or thousands intermediate operations executed during object access. All these operations are executed transparently behind the scenes but the programmer has complete control over these levels of the program.

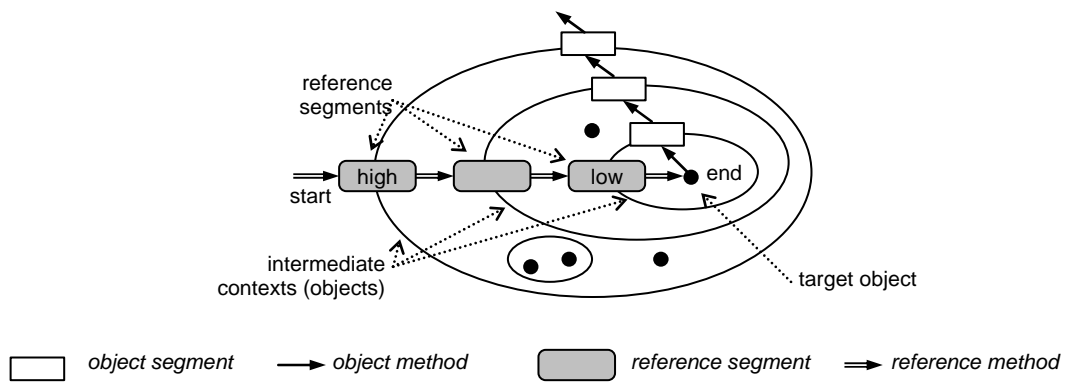

Figure 18. Computation is a process of border intersection during object access.

Informally, the classical approach to programming is analogous to the action-at-a-distance principle in classical physics dominating in 18[th] century. In contrast, the concept-oriented paradigm assumes that objects cannot interact instantaneously because any interaction needs some environment to propagate. When we want to access an object this environment will get control over the process of propagating the access request to the target object. In the proposed approach it is postulated that this logic of propagation of interactions cannot be avoided and hence any access is by definition *indirect*. In other words, any access needs some time and intermediate processing to be performed. Any reference in this case is a convention and it needs to be interpreted by the layer that issued it. References are used to describe where an object resides while the space where objects live mediates their interactions. The space or environment in this case plays an active role and intervenes in the process of object interactions. Depending on the space geometry an object may behave differently. In other words, if one and the same object is placed in one container then it will exhibit one type of behaviour but if the container changes then this object looks differently. Objects are an entity in itself and can be viewed only via the prism of their references and the space they live in. Here we can remember a principle of the relativity theory which says that objects in the space change its geometry and the geometry influences the objects that exist in it. In CoP the situation is the same: objects are not simply end points but rather implement functions of space with some specific geometry and influencing other objects.

Such a system can be represented as a structured space where elements are separated by the space borders. In this case an interaction is only possible by intersecting the borders. Hence it is frequently more important what happens during access than in the end point. The process itself tends to be more important than the goal. Such behaviour is very characteristic for complex systems where a great deal or even most of the functionality is concentrated on space borders rather than in end points. Thus we change the traditional view how functionality is distributed over the system elements. If in OOP it is assumed to be encapsulated in objects then in CoP it is assumed that functionality is encapsulated in intermediate elements. In fact, we cannot say where one or another function will be executed because each intermediate element will make its own contribution. Concept as a programming construct has been precisely adapted to such a view. In particular, concept methods (for both reference class and object class) should be thought of as intermediate processing points which are called from somewhere and then pass control further to somewhere. The programmer describes what a method has to do *if* it gets control being unaware from where concretely it gets it and how it will proceed further. (The difference between reference methods and objects method is that they have opposite continuation direction.)

In the concept-oriented paradigm computations are thought of as a path crossing the space borders and then the question is how this space can be modelled. And here again concept appears to be an adequate tool which allows us to describe a hierarchical space structure using inclusion relation. References and objects are living within one hierarchy playing however the dual roles. An object may have many child objects (extensions) for which it is a context. Thus in contrast to OOP, objects may have many extensions distinguished by their local references. The role of references consists not only in representing objects but also in protecting object methods from direct execution. So reference is a front end to its object and to the internal space consisting of child references and objects.

Program development in OOP focuses on describing object behaviour via classes what can be expressed as a slogan: "Describe your classes and you will get your system". In the concept-oriented paradigm the focus shifts in the direction of references and one of the main concerns consists in describing their layered structure. This structure is actually a container for all objects of the program with its own virtual address system. Defining the virtual address system is what the concept-oriented development process should start because objects cannot live in vacuum. They cannot be homeless and need some environment providing all the necessary facilities like life-cycle management. This principle could be expressed as the following slogan: "Describe a virtual address system for your objects and you will get half of the system".

The virtualization is very important term here because in the concept-oriented approach we are not bound to any reality when developing a system. When in OOP we are talking about objects then we explicitly or implicitly assume some memory. In CoP objects are living within a *virtual* address space which is being developed by the programmer. For example, we can define references are texts strings or integers and then each object will have such an identifier. There is no indication that these objects reside in memory, on disk, on tape or on any other concrete device. We have an identifier from the virtual address space and that is all what we have and actually that is all what we need to work with objects. All the connections with the reality are implemented at lower levels of the system organization in base concepts.

In such a virtual address system an object lives in a complex environment which provides many important functions like protection and life-cycle management. Thus objects not only get useful services but also depend on the environment. In particular, an object may require certain services and hence can live only in some type of environment (requiring some type of parent concept in CoP). On the other hand, environment itself may contain only objects of certain type which obey special laws. But what is even more important, the behaviour of objects can depend significantly on the container where it is currently in. An object method may have one result when the object is in one container and another result when this object is put in another container. Such a context dependent behaviour implemented by a container is also known as *dependency injection*.

CoP can be viewed as an approach to describing cross-cutting behaviour as understood in AOP. Indeed, the behaviour encapsulated in parent concepts is effectively applied to the requests passed to the child objects. Parent concept functions intercept calls to child functions and hence they can intervene whenever a process intersects the border on its way to a destination within this context. However, CoP and AOP use not only different technical means and language constructs but also are based on very different general assumptions on how a system is organized and how it functions so their similarity is rather shallow. One of the most interesting common properties of the both paradigms is that there is some intermediate behaviour executed automatically behind the scenes.

# 11  Conclusion

In the paper we described a new approach to programming which generalizes object-oriented programming. The main programming construct in this approach is called concept and consists of two classes: one reference class and one object class. Concepts are used instead of classes for declaring type of elements of the program such as variables, fields, parameters and return values. The main advantage of the described approach is that it allows the programmer to describe not only functions of objects which are used explicitly, but also functions of references which are used implicitly.

One direction for future work consists in establishing closer connections with the concept-oriented data model [Sav05, Sav06, Sav07]. The thing is that principles of CoP described in this paper are analogous to the physical structure of the concept-oriented data model (CoM). In particular, CoP is precisely what is called identity modelling in CoM. The techniques for modelling format and functions of references can be used for modelling identifiers of entities in data modelling.